# A catalog of old globes in Spanish public collections

**Miguel Querejeta**
Observatorio Astronómico Nacional
Madrid, Spain

## Abstract
This paper presents the first catalog of celestial and terrestrial globes, as well as armillary spheres and orreries, produced before 1900 and preserved in Spanish public institutions. Most globes have an English or French origin, predominantly from the late 18th or 19th centuries. We highlight a few outstanding examples, including an early metallic terrestrial globe, a mysterious blue celestial manuscript globe, the oldest preserved Spanish printed globe, and some interesting clockwork pieces. While Spain has not been a major producer of globes, it does preserve around two hundred historical globes in public collections, including several remarkable pieces.

### Introduction

Terrestrial and celestial globes are spherical representations of the Earth and heavens, and they constitute a prime example of the connection between cartography, astronomy, and art. Throughout history, globes have been regarded as a symbol of knowledge and status, and they have played a key role as scientific instruments with useful applications to navigation.[1] The idea of producing spherical models of our planet and the heavens dates back, at least, to classical Greece. We find an outstanding example of this in the Farnese Atlas, a marble sculpture currently on display at the National Archaeological Museum of Naples. The titan Atlas holds a large sphere with beautifully delineated constellations, many of which were compiled in Antiquity and are still in use nowadays.[2]

A good number of celestial globes were engraved onto metallic spheres during the Middle Ages, primarily associated with the Arab world.[3] However, it was with the development of printed globes, starting from the 16th century, that their production became much more widespread.[4] Globes were typically printed on a set of elongated gores that would then be glued onto a sphere made of papier-mâché and plaster. This is the kind of globe that highly successful Dutch manufacturers such as Mercator, Hondius, Blaeu, or Valk developed, and which feature in paintings such as Vermeer's famous *Astronomer*.[5] Many globe producers established family businesses that continued across generations.[6] The development of printed globes allowed for multiple copies of the same globe to be made, but, at the same time, the fragility of the plaster sphere implies that many have not survived.

For over three centuries, starting from the 16th century, globes were often produced in pairs: a celestial and a terrestrial globe, both of equal diameter, intended to be displayed side by side.[7] This way, globes represented the ultimate synthesis of cosmography, the union of cartography and astronomy. They were sometimes complemented by armillary spheres, which are idealized representations of the orbits in the solar system made up of rings.[8] From the 18th century onwards, sophisticated mechanical devices such as orreries were also developed, illustrating the motions of the solar system. It was common to find globes in libraries, and also in wealthy households, observatories, schools, and universities.

Building upon the seminal work by Stevenson,[9] catalogs of old globes have been published for a large number of countries in Europe, for Japan and the United States.[10] Such inventories constitute an invaluable tool for researchers, librarians, curators, and even private collectors. However, a catalog of old globes preserved in Spain was still lacking.[11] Some publications have examined specific globes in Spain, such as those by the Blaeu family,[12] and a number of papers report on the restoration of individual globes, including a pair by John and William Cary in Tenerife,[13] another pair by Dudley Adams at Museo de América,[14] or a celestial globe by Dudley Adams at Universidad de Salamanca.[15] Hernando[16] traced back the production of globes in Spain on written records, concluding that globes were produced in the Iberian Peninsula as early as the Middle Ages, but most of them seem to be lost. Regarding more recent Spanish producers, Hernando discusses Tomás López, Antonio Monfort, and Faustino Paluzie, who are all present in Spanish collections, as will be shown below.

In this work, we present a catalog of globes, armillary spheres, and orreries produced before 1900 and held in Spanish public collections. Our goal is to identify which kind of globes exist in Spain, paying attention to their epoch, type of globe, and geographical origin. This paper is structured as follows. We start by outlining the methodology and criteria for a given piece to be included in the catalog. Then, we comment on some of the earliest globes and armillary spheres held in Spain, and present some statistics on the properties and geographical distribution of the pieces from the catalog. We devote specific sections to some outstanding globes from Spanish public collections. We close the paper with a summary and conclusions.

## Methodology

In order to compile a catalog of old globes in Spanish public collections, we carried out an extensive search across a wide range of institutions. Firstly, we contacted the institutions associated with the IBERCARTO project,[17] which include the most relevant cartographic collections in Spain. We also reached out to the different sites of Patrimonio Nacional, which preserve important collections from the Spanish monarchy that now belong to the State. Additionally, we contacted an extensive list of museums, identifying them through the directory of national, regional, and local museums listed under the website of the Ministry of Culture.[18] We also approached university libraries and heritage departments from universities, selecting those founded before 1900. We also included historical high schools (*institutos históricos*), complemented by an extensive online search for hints of additional globes in public collections such as observatories, public foundations, and historical institutions such as *Ateneos* and *Sociedades de Amigos del País*.

We limit the time frame of the search to globes produced before 1900; most catalogs of old globes stop either at 1850 or 1900, so this is a reasonable time frame. Following the example of other published globe catalogs, we consider three-dimensional objects encompassing celestial and terrestrial globes, armillary spheres, planetaria, and orreries,[19] as well as rare pieces such as tellurions[20] and moon globes. We also consider globes which are part of clocks or other pieces such as decorative silver objects, as long as the globes are a detailed cartographic representation of the earth or the heavens (i.e. excluding those which are merely ornamental and symbolic). We also take into consideration astronomical clocks when they include the equivalent of an orrery, but we deliberately exclude globe gores and atlases, which are sometimes included in these kinds of catalogs.

For each object, we collected the following information: institution where the object is held, type of object (celestial globe, terrestrial globe, etc.), manufacturer, year, and place of production (when known), the diameter of the sphere, and its inventory number if applicable. In some cases, this information was readily available from museum or library catalogs, but in most cases we had to examine the globes, carry out measurements and infer the date of production. The diameter refers to the sphere itself, excluding any meridian or horizon rings; we derived it by measuring the equatorial perimeter with a string and dividing it by π. The year of production is often provided in a cartouche that also lists the manufacturer; when this piece of information is lacking, an approximate date was estimated. For terrestrial globes, the date can be inferred from the details of political borders and toponymy; for celestial globes, the presence of certain constellations and the precession of equinoxes permits to derive an approximate dating by studying the exact point where the ecliptic crosses the celestial equator. These inferred dates are indicated as *ca.* in the catalog. Even for globes with an explicit release date, the same globe often continued to be produced without change for some years afterward. In the comments section of Table 2, we also highlight manuscript globes and point to relevant references.

## Early globes in Spanish public collections

Table 2 presents our main output, the catalog of historical globes in Spanish public collections. It contains a total of 199 pieces: 57 celestial globes, 82 terrestrial globes, 48 armillary spheres, tellurions or planetaria, 11 orreries, and a lunar globe. This inventory identified 21 pieces (11%) produced before 1700. The oldest globe in the catalog might date back to the beginning of the 16th century, which would qualify it as one of the oldest metallic terrestrial globes in the world; we examine this "globe of Bailly" in more detail below. One of the best-known cartographers of the 16th century, Gerard Mercator, is also represented in the Spanish public collections with a pair of 42 cm celestial and terrestrial globes at the National Archaeological Museum.[21] As studied in detail by Martín-Merás,[22] Spain preserves a good number of globes from the Blaeu family, including four pairs of 68 cm globes (at Museo Arqueológico Nacional, Escorial, Salamanca, and Valencia), a smaller 34 cm pair at Real Academia de la Historia, and two individual celestial globes, both of 34 cm in diameter, one at Museo Arqueológico Nacional and the other at Museo Nacional de Ciencia y Tecnología.[23] In the Museo Nacional de Ciencia y Tecnología we also find one of the earliest pocket globes, produced by Joseph Moxon in England in the second half of the 17th century. This small terrestrial globe (7 cm) has a case made up of two concave hemispheres depicting the heavens; Moxon is indeed sometimes quoted as the inventor of the pocket globe.[24] Finally, two precious globes by Vincenzo Coronelli are displayed at Museo Naval in Madrid; this pair of terrestrial and celestial globes, 108 cm in diameter, were produced in 1688 and 1693, respectively.

Apart from globes, some of the earliest instances in the catalog correspond to armillary spheres. The most remarkable early example is a large armillary sphere at Escorial which is primarily made of wood. The astronomer and cosmographer Antonio Santucci built it in 1582 for cardinal Ferdinando I de' Medici (who would become Grand Duke of Tuscany); his coat of arms is visible on the terrestrial globe at the center of the sphere. The armillary sphere was given as a diplomatic gift to King Philip II of Spain back in 1582, and he placed it in the library of the monastery of El Escorial in 1593. Santucci would also build a similar, larger armillary sphere of roughly two meters in diameter in 1588-1593 for the Sala delle Matematiche in the Uffizi, now on display at Museo Galileo in Florence. Also in the library at Escorial, a smaller armillary sphere made of brass attributed to Michel Coignet likely dates back to the late 16th century. A similar armillary sphere of brass signed by Coignet, dated 1591, is held at the Adler Planetarium in Chicago.[25] Finally, an anonymous armillary sphere today at Museo Nacional de Ciencia y Tecnología may have been produced in the 17th century.

In addition to these early globes, the Spanish public collections include some exceptional pieces from the 18th and 19th centuries, among them a mysterious blue celestial globe and a pair of armillary spheres from the late 18th century, the earliest extant terrestrial globe produced in Spain, as well as some remarkable orreries. We examine them in detail below.

## An inventory of globes in Spain: statistics on types, origin, and location

The earliest pieces in the catalog show a higher proportion of celestial globes, and more generally of globe pairs. Out of the 199 globes from the catalog, 54 globes form celestial-terrestrial pairs by the same manufacturer and of the same diameter (i.e. 27 pairs).[26] The percentage of globes in pairs is 62% for globes before 1700, 33% for globes from the 18th century, and 21% for globes from the 19th century. Figure 1 shows the evolution in the percentage of the different types of globes over time in our catalog. While celestial globes represent 38% of the pieces before 1700, this fraction slowly declines throughout the 18th century and drops down to 24% in the 19th century, when terrestrial globes dominate, particularly towards the end of the century.

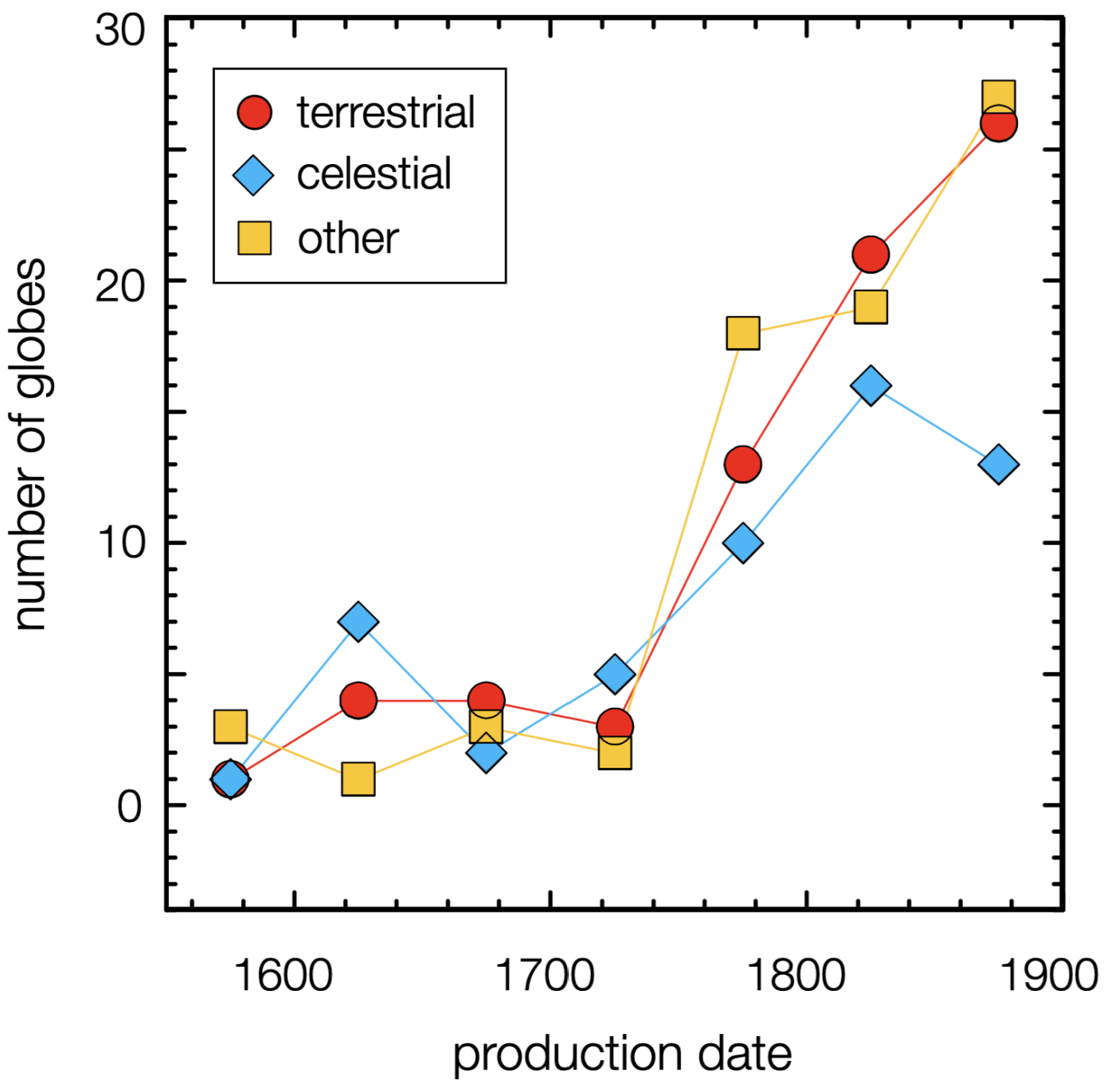

**Fig. 1**: Distribution of globe types over time in Spanish public collections.

Regarding the origin of globes in Spanish public collections, as can be seen in Table 1, France is the country with the largest overall percentage of globes in the catalog (36%), followed by England (19%), Spain (15%), and the Netherlands (9%). The notorious presence of French globes is mostly due to the explosion of the import of French globes in the 19th century (when French globes account for as much as 40% of the total). On the other hand, if we pay attention to the earliest globes (before 1700), the Netherlands holds the record with 62%, primarily as a result of the good number of globes from the Blaeu family in Spain. In the 18th century, on the other hand, English globes were the most popular (39%), closely followed by French globes (35%).

| | Until 1700 | 18th century | 19th century | Total |
|---|---|---|---|---|
| **France** | 4 (15%) | 18 (35%) | 49 (40%) | 71 (36%) |
| **England** | 1 (4%) | 20 (39%) | 17 (14%) | 38 (19%) |
| **Spain** | 0 (0%) | 6 (12%) | 23 (19%) | 29 (15%) |
| **The Netherlands** | 16 (62%) | 1 (2%) | 0 (0%) | 17 (9%) |
| **Others/unknown** | 5 (19%) | 6 (12%) | 33 (27%) | 44 (22%) |
| **Total** | 26 | 51 | 122 | 199 |

**Table 1**: Number of globes sorted by country of origin and production period. Percentages indicate the proportion of globes corresponding to each country in each time interval.

We also examine the distribution of globes geographically within Spain. Figure 2 shows the distribution of globes per autonomous region. Madrid stands out as the region with the highest number of globes, with as many as 95 (roughly 50% of the catalog). This is not surprising given the fact that, as the capital of Spain, most globes associated with the crown (Escorial, Palacio Real, etc.) are in Madrid, as well as globes belonging to national institutions, such as the National Library, National Archaeological Museum, or Museo Naval. The second region in number of globes is Catalonia, with 22 globes (11%), probably due to the historical relevance of Barcelona. This is closely followed by Galicia (10%), which could be explained by the connection with navigation, and Castilla y León (9%), where the historical universities of Salamanca and Valladolid account for most historical globes.

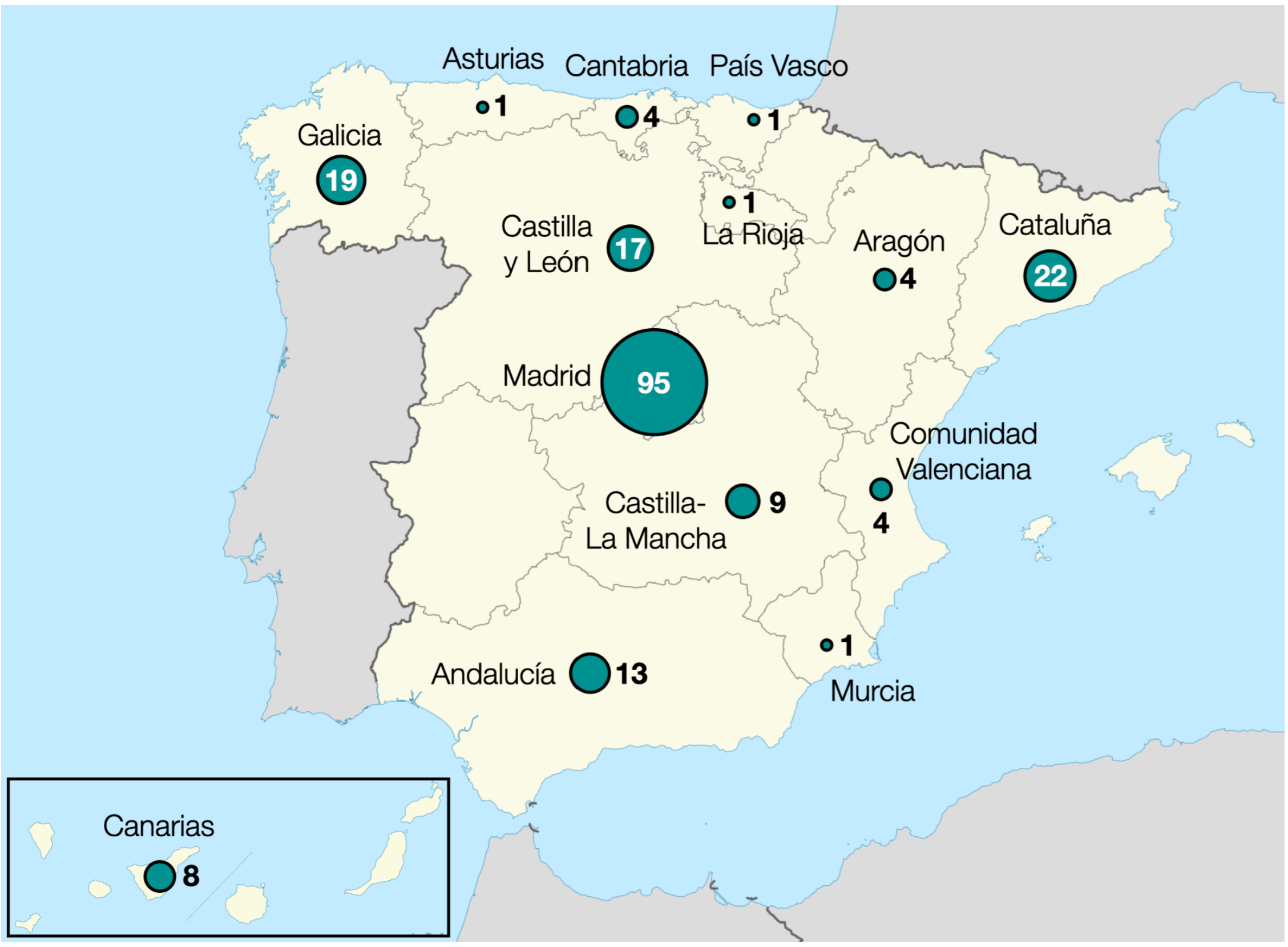

**Fig. 2**: Geographical distribution of historical globes in Spanish public collections.

Regarding the type of institution where the globes are currently preserved, 54% of the pieces in the catalog are in museums (including the sites of Patrimonio Nacional), 33% are in libraries, universities, or research institutes, while 9% are found in historical schools. Figure 3 shows how the relative distribution of globes among the different kinds of institutions has evolved over time (as a function of when the globes were produced). The proportion of older globes in museums is comparatively higher, while in the 19th century there is a trend for more globes to be found in libraries or research institutions. Towards the end of that century, there was a pronounced uprise in the fraction of globes at schools (which is not too surprising, as quite a few of these *institutos históricos* were founded in the second half of the 19th century).

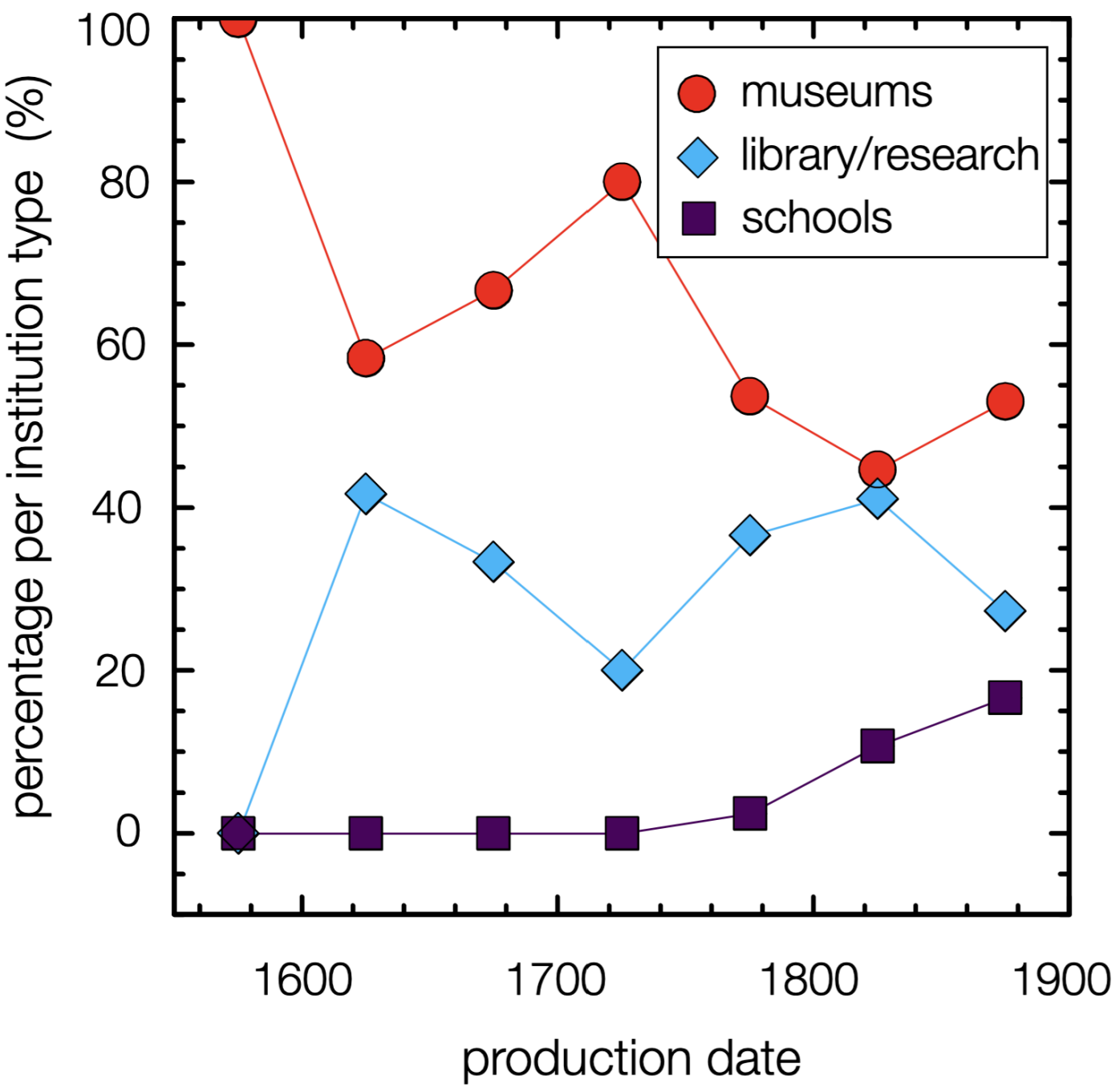

**Fig. 3**: Evolution over time in the type of institutions where globes are held.

## An early metallic globe: the globe of Bailly

The Lázaro Galdiano Museum in Madrid houses the private collection of José Lázaro Galdiano, passed on to the Spanish state upon his death in 1947. A piece of gilded silver is particularly relevant to this work, as it contains a terrestrial globe that may have been produced in 1530. The small metallic terrestrial globe of 14 cm in diameter sits on top of an Atlas-like figure, and opens in two halves, which qualifies it as a cup (Fig. 4). The object once belonged to the collection of Carl Mayer von Rothschild and was acquired by José Lázaro in New York in the 1940s.

The cartography of America, the southern part of Africa, and Asia closely resembles the globe now held at the Morgan Library in New York,[27] purchased by J. Pierpont Morgan from Ludwig Rosenthal in 1912. The latter is signed "ROBERTVS DEBAILLY 1530" and coincides in diameter with the globe in Madrid (14 cm).[28] In fact, the globe from Madrid has a virtually identical signature in the same position on Antarctica, which reads "ROBERIVS [sic] DEBAILLY 1630". The public catalog from Lázaro Galdiano interprets the year 1630 as a typo that should read 1530, following Siebold,[29] who quotes the globe of Madrid as a second copy of the globe from New York. However, a close examination of the globe of Madrid shows that the inscription 1630 appears twice in Antarctica (while the globe from the Morgan Library reads 1530 exactly in the same positions for those two instances), and it seems hard to believe that the same typographic error appeared twice.

The Lázaro Galdiano globe is plagued with misspellings in the toponymy compared to the globe of New York (CIRCVLUS ARTICVS > CIRCVLUS ARTRCVS; VERRAZANA > ERRA□ANA; S. MICHAELIS > MIHAELIS; R. DVLCE > R DVIGE; MARE INDICVM > MARNEINDICVM; INSVLA S. LAVRENTII > INSVLAS AVRENTII; SOGDIANA > SCGDKANA; TARTARIA > TARTAIRNE; DESERTV > DERERTV...). Most strikingly, while the cartography of America, southern Africa, and Asia closely follows the model of the globe from the Morgan Library, in Europe and northern Africa the cartography seems far more archaic and inaccurate, to the point that well-known geographical features such as the Italian peninsula are essentially lacking (Fig. 5). These differences and the multiple misspellings in the toponymy suggest that the globe of Madrid was not made by the same cartographer as the globe of Bailly; rather, it seems likely that it was engraved by an artisan who was illiterate or, at least, not very competent in Latin and not familiar with cartography. However, beyond any doubt, there is a close connection between the manuscript terrestrial globe from Madrid and the globe of Bailly. The choice of toponyms, including "VERRAZZANA", despite the linguistic corruptions, and even the positioning and design of decorative drawings such as vessels or animals (e.g. exactly the same elephant in Africa, Fig. 5), strongly suggest that the globe of Madrid was partially copied from the globe of Bailly (or from a common source). While it is possible that the globe was copied in 1530 or shortly after, the inscription of 1630 twice would rather indicate that this version dates back to the first half of the 17th century. This dating of 1630 for the globe was accepted by Camón Aznar[30] and Cruz Valdovinos,[31] who argued for a German or Austrian origin and pointed out that the Atlas-like sculpture and the naked boy looking through a telescope on top would likely also date from around that time. Indeed, the telescope became very popular in all kinds of artworks after the discoveries of Galileo in 1609,[32] so this sounds plausible. Yet, the reason why the cartography of Europe and northern Africa does not at all follow the model of the globe of Bailly remains a mystery.

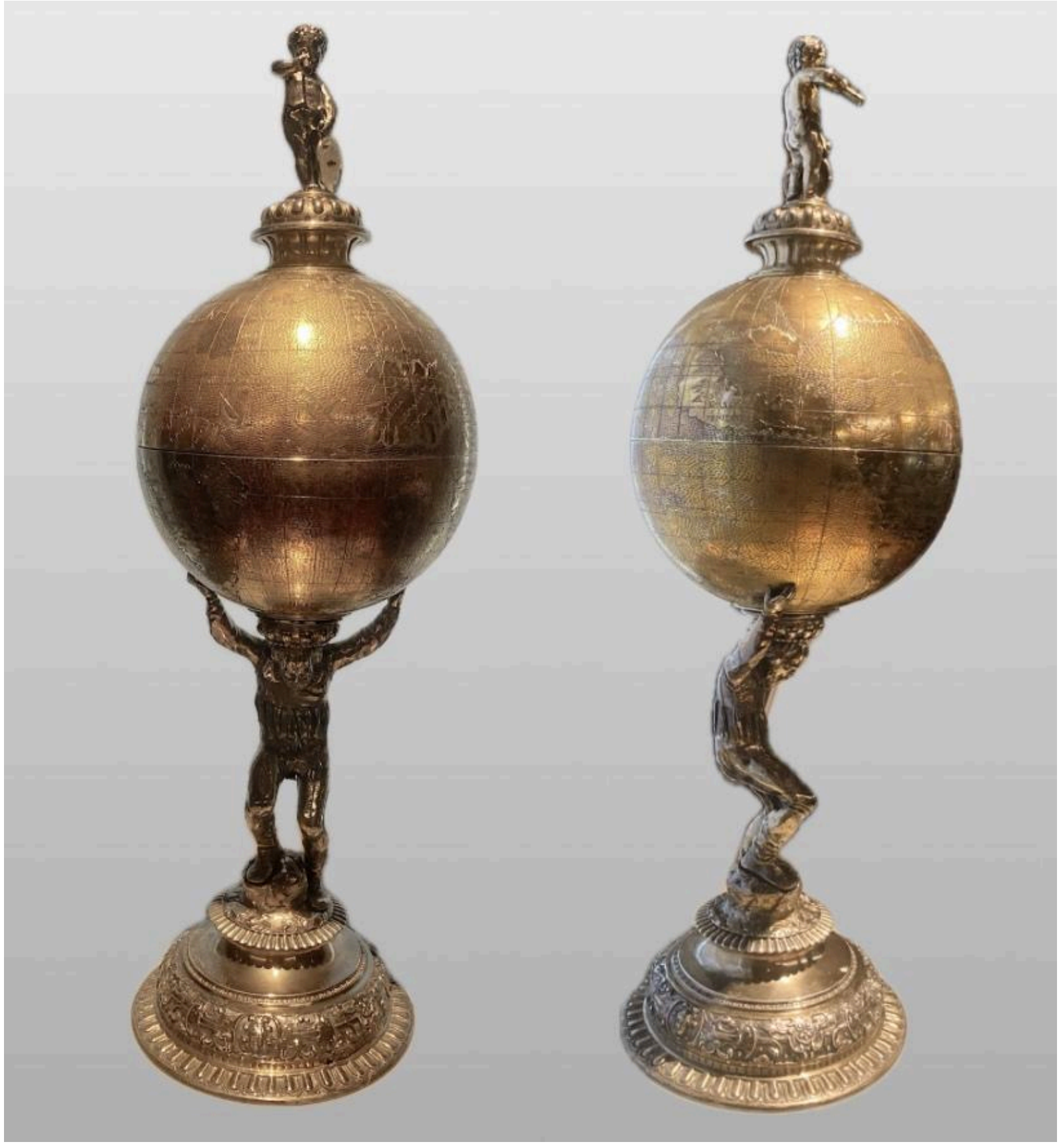

**Fig. 4**: The so-called "globe of Bailly" at the Lázaro Galdiano Museum in Madrid.

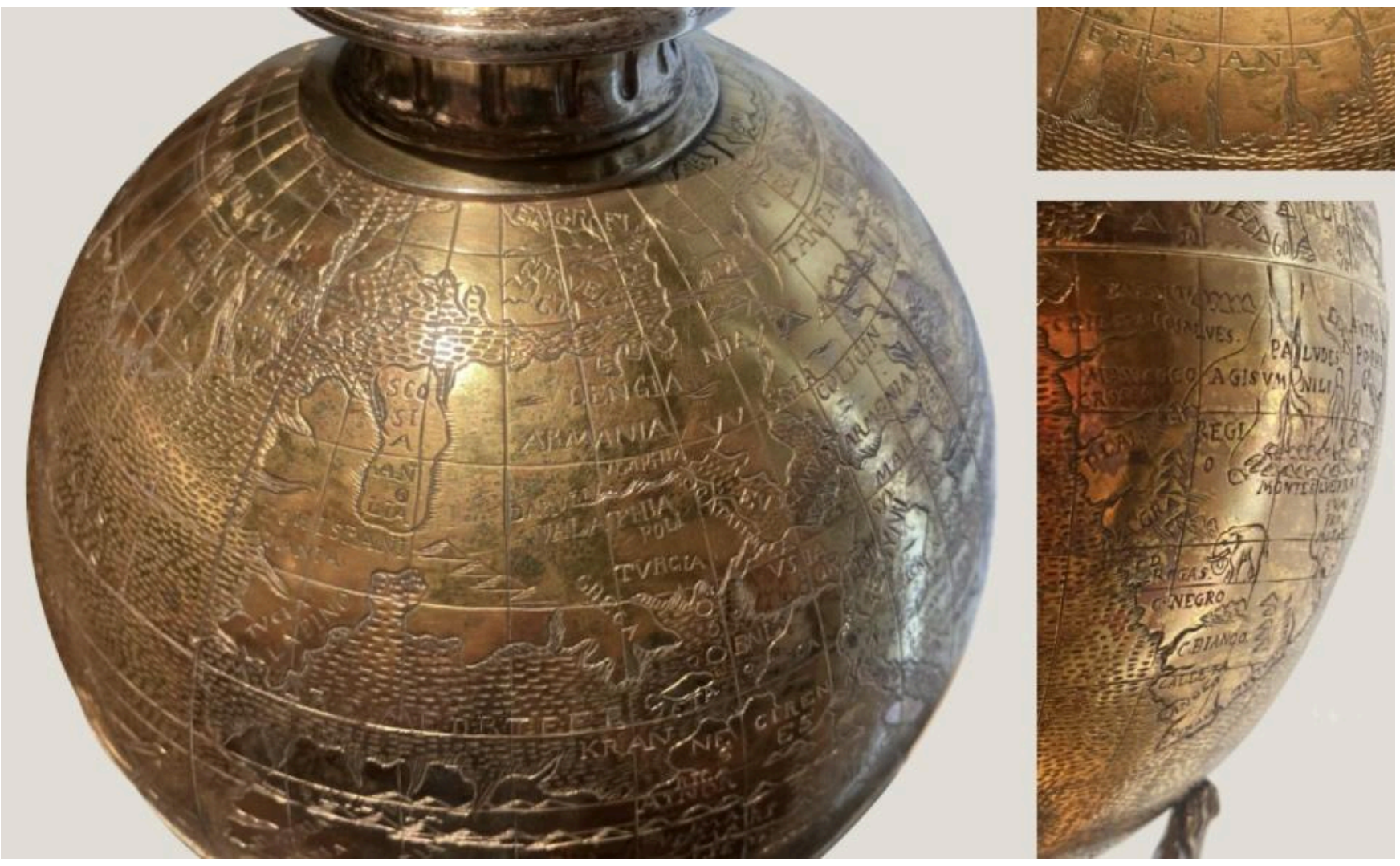

**Fig. 5**: Details of the "globe of Bailly" at the Lázaro Galdiano Museum in Madrid.

## The oldest terrestrial globe produced in Spain: Tomás López

Several written records point to globes produced in Spain at least since the Middle Ages.[33] However, according to our catalog, the earliest Spanish globe still preserved in public collections of this country is the 18th-century terrestrial sphere by Tomás López (Fig. 6). Currently held at the National Library of Spain, it was purchased in 2010 and it came from the private collection of an Italian aristocrat. It is a small globe made of papier-mâché and plaster on top of which 12 engraved paper gores plus two polar caps were glued; the support, which seems original, is made of wood. The library catalog, Líter Mayayo,[34] and Hernando[35] quote a diameter of 26 cm, but direct measurements carried out for this work yield a slightly larger diameter of 30 cm for the sphere.

While the globe is signed, there is no explicit indication of the year when it was produced. As pointed out by Líter Mayayo,[36] the globe must have been produced after 1770, when Tomás López was appointed King's Geographer, because this title is listed in the cartouche of the globe ("Por D. Tomás López, Geógrafo de los Dominios de S. M."). Líter Mayayo further argues that the globe must have been produced before 1792, as a world map by Tomás López from that year already shows the eastern coast of Australia outlined. Indeed, a close inspection of the globe confirms that it originally left the eastern coastline of Australia open, and it was later retouched, possibly in pencil, to close the outline of Australia once it was confirmed to be an island. Here we argue for an even narrower dating window between 1770 and 1780. Between 1770 and 1792, Tomás López, an extremely prolific cartographer, produced a large number of maps of Spain, including detailed regional maps. The Spanish coastline of the globe agrees better with the earlier maps (of around 1770-1780), especially for the southern tip of Cádiz as well as the shape and relative positioning of the Balearic islands. However, an even more compelling argument that we highlight here is an inscription found in front of present-day northwestern Alaska, which reads "…. de 30 años a esta parte" ("in the last 30 years"). This is essentially the same text present in the world map of 1771 by Tomás López,[37] alluding to the islands discovered by Russian explorers between 1740 and 1770. If the globe were produced close to 1790, it seems hard to believe that Tomás López did not update this note, since he had to produce the plates specifically for the globe (in fact, he made a minute modification to the text, from "treinta años" to "30 años").

To our knowledge, this is the only globe by Tomás López known in the world. This might be somewhat surprising, given that he was the most important Spanish cartographer of the 18th century, a time when globes were very fashionable; in fact, according to the inventory of his private library, published by Patier Torres,[38] he possessed "several spheres for producing globes, including the copper plates, 6" ("Varias bolas para globos, con los cobres, 6"). Interestingly, a few globes by Pedro Martín de López, who married the niece of Tomás López's son and considered himself the successor of the family business, are known.[39] These globes have the peculiarity that they are foldable.

Summing up, the terrestrial globe by Tomás López is the oldest terrestrial globe produced in Spain which is still preserved in Spanish public collections. This demonstrates that, similarly to other cartographers of the time, not only did Tomás López produce a prolific collection of maps, but also terrestrial globes. As we have argued here, the globe must have been produced with almost absolute certainty between 1770 and 1792, and most likely between 1770 and 1780.

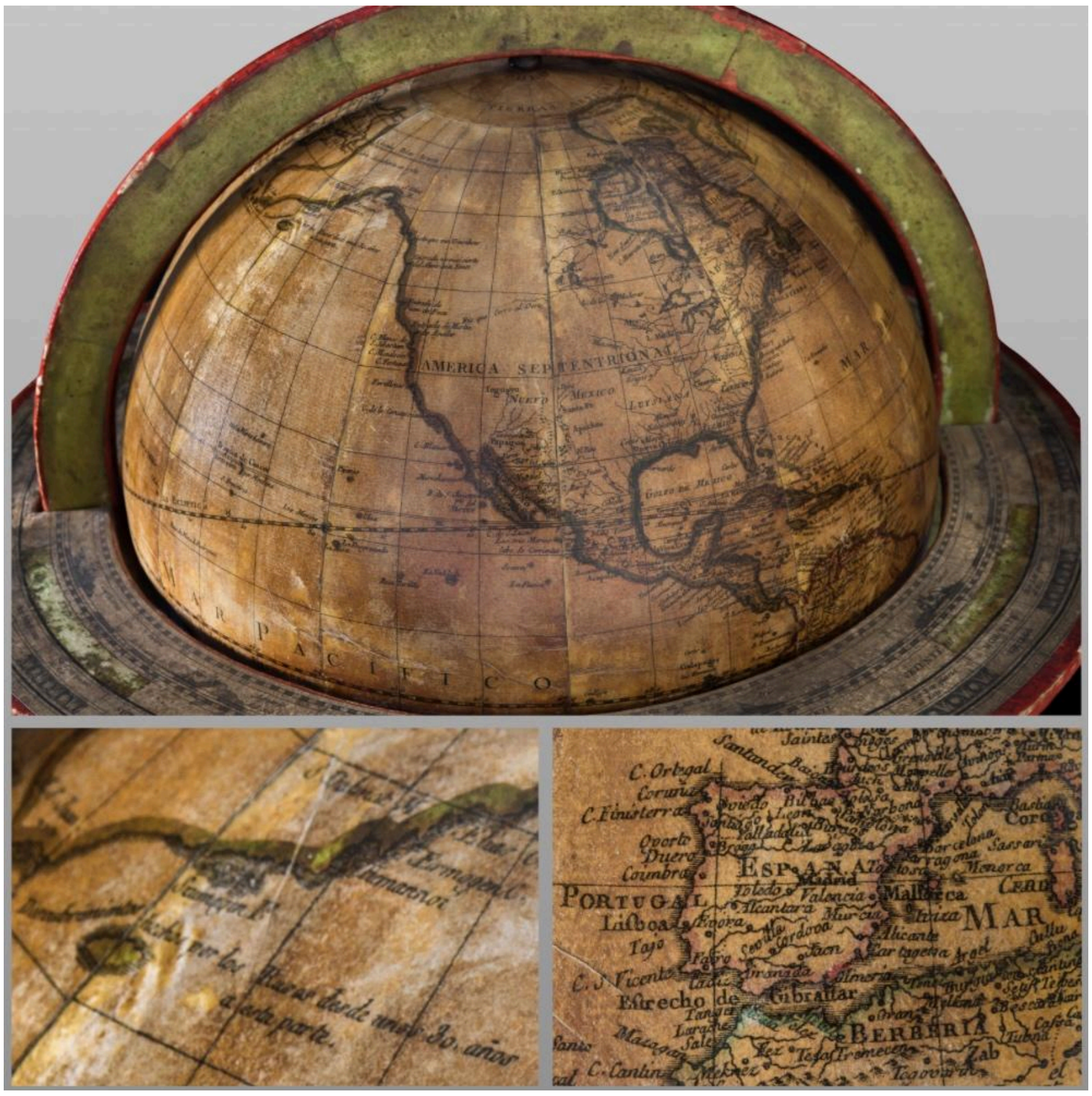

**Fig. 6**: Details of the terrestrial globe by Tomás López (Biblioteca Nacional de España, Madrid). The bottom-left panel shows an inscription that supports a production date around 1770-1780.

## A mysterious celestial globe and two armillary spheres

The Spanish National Library preserves a remarkable 68 cm blue celestial globe of unknown origin. The support of the globe suggests that it dates back to the 18th century.[40] However, the producer or even country of origin remains a mystery. The celestial globe shows a dark blue background with constellation drawings in gold. The stars are metallic cutouts, mostly golden and sometimes reddish, pasted on the sphere; they range in size and number of vertices, roughly reflecting magnitudes. Restored in 1997, the photographs taken prior to the restoration[41] show that the globe was in a relatively poor state; many metallic stars had fallen off and were reintegrated during the restoration process.

One of the most unusual features of the globe is that it lacks any cartouches or indications of authorship, which is extremely uncommon for globes before 1800. Even though a quick look conveys the impression that the globe lacks any annotations, the names of the constellations and months are indicated in relief in Latin, in the same blue hue as the background. The dark blue color departs from the 18th century tradition of printed globes, which typically have light backgrounds.

Indeed, according to our catalog, all celestial globes older than 1850 preserved in Spain display a light background. The same applies to the vast majority of globes from this time abroad. The few cases of blue globes before 1850 are exceptional examples of manuscript globes, often presents for kings or highly distinguished people, such as the famous pair of globes produced by Vincenzo Coronelli for King Louis XIV (Bibliothèque national de France),[42] the beautiful 16th century celestial globe by Heinrich Arboreus (Bayerische Staatsbibliothek, Munich),[43] or the pair of globes by Mercator which were presumably a present for the Spanish emperor Charles V (Vienna Globe Museum).[44] The blue-green celestial globe by abbé Nollet, today at Bibliothèque nationale de France, also constitutes an interesting exception. As indicated by Hofmann,[45] it is not only the blue color of that early 18th century globe that makes it special, but also the fact that the constellation drawings are delineated in a very subtle way, as it would become common after 1850, to convey a more realistic impression compared to the actual night sky. Another analogy with the blue globe from the Spanish National Library is the fact that stars are golden cutouts which vary in size and number of vertices.

Our work has revealed a connection between the blue celestial globe from the Spanish National Library and a pair of globes by Dudley Adams currently held at Museo de América in Madrid.[46] They have exactly the same diameter[47] and a nearly identical wooden stand (see Fig. 7). The remarkable similarities between the globe of Dudley Adams and the blue celestial globe in terms of size and supporting structure, as well as the design of the constellation drawings, suggest that there might be a close connection between them. While it is possible that the blue globe was produced by Dudley Adams, to our knowledge, no other blue globes by Dudley Adams are preserved anywhere else in the world and, furthermore, it seems unlikely that a professional manufacturer of globes such as Adams would not sign such a globe.

The blue celestial globe from the Spanish National Library has been nicknamed "the globe of Godoy" after Manuel Godoy, the Spanish prime minister under King Charles IV (1792-1808).[48] With the onset of the Peninsular War in 1808, both Charles IV and Godoy had to flee Spain, and the belongings of the prime minister were largely plundered. Part of his library, presumably including this remarkable celestial globe, passed over to the National Library, where it has been until today. This seems plausible, particularly considering the inclination of Godoy to collect scientific instruments.[49]

Here we propose the hypothesis that the blue celestial globe was produced at the Royal Observatory of Madrid, at the newly established workshop of instruments, in the decade of 1790. Carlos Rodríguez and Mario Fernández, two young Spanish astronomers, were actively trained in London around 1790 in the art of producing astronomical instruments;[50] the contact with Dudley Adams seems highly likely given the coincidence in space and time. The similarity between a telescope by Carlos Rodríguez and Mario Fernández (dated 1790, Fig. 8) and a telescope by Dudley Adams also supports the idea that there was a probable connection between them.

The main motivation behind our hypothesis that the blue globe was produced in Madrid is a piece of text written by Antonio Gil de Zárate in the mid-19th century, which tells us that, by 1795, the director of the Royal Observatory of Madrid gave two globes as presents to the king, Charles IV, produced at the workshop of the Observatory. If the king received such a blue celestial globe, it seems plausible that it passed over to Manuel Godoy, who was passionate about scientific instruments and called himself "protector of the Observatory" according to his own memoirs;[51] the globe would have ended up at the National Library together with his private library. Given the likely connection between Carlos Rodríguez and Mario Fernández and Dudley Adams, we speculate that they could well have used some printed gores by Adams as a template for their manuscript globe (which would explain the coincidence in diameter), and have copied the piece of furniture that supports it, or even have brought it over from England to Madrid. It is also possible that they repainted the globe in blue on top of an original, possibly damaged globe by Dudley Adams. In any case, the choice of a blue color could be an attempt to make it special as a gift for the king, since Charles IV already had some globes purchased abroad,[52] including several globes by George Adams (father of Dudley Adams).

This brings us to two other interesting pieces that stand out in the catalog: a pair of armillary spheres (geocentric and heliocentric) made of brass. They are held at the Royal Observatory of Madrid, and have traditionally been attributed to the workshop of instruments that was active at the Observatory between 1794 and 1808.[53] The armillary spheres are not signed, but they contain inscriptions in French (planet names, months, zodiac signs). This could seem surprising if the spheres were produced in Madrid, but it might be explained by the fact that the director of the workshop of instruments, Pierre Mégnié, was French.[54] The similarity in material and aesthetics with a Gregorian telescope signed by Carlos Rodríguez and Mario Fernández (Fig. 8) supports the plausibility of the attribution to the workshop of instruments. In fact, this type of foot is uncommon for armillary spheres, and resembles a telescope foot.[55]

The vast majority of armillary spheres in the catalog are made primarily of wood, mostly from the 19th century. Coeval with the pair of armillary spheres from the Royal Observatory of Madrid, several armillary spheres by members of the Rostriaga family were also produced in Spain and are made of brass.[56] However, their design closely follows the brass armillary spheres by George Adams,[57] while the spheres from the Royal Observatory of Madrid depart from this tradition.

In any case, we are dealing with two armillary spheres of outstanding beauty which most likely date back to the late 18th or early 19th century. According to our hypothesis, these armillary spheres, together with the blue celestial globe, could be the legacy from the ephemeral workshop of instruments that existed between 1794 and 1808 at the Royal Observatory of Madrid, which at that point was at the forefront of astronomical research. The Observatory had purchased a 25-foot telescope from William Herschel and two young astronomers were trained in London over the course of five years, where it is highly probable that they were in contact with Dudley Adams.

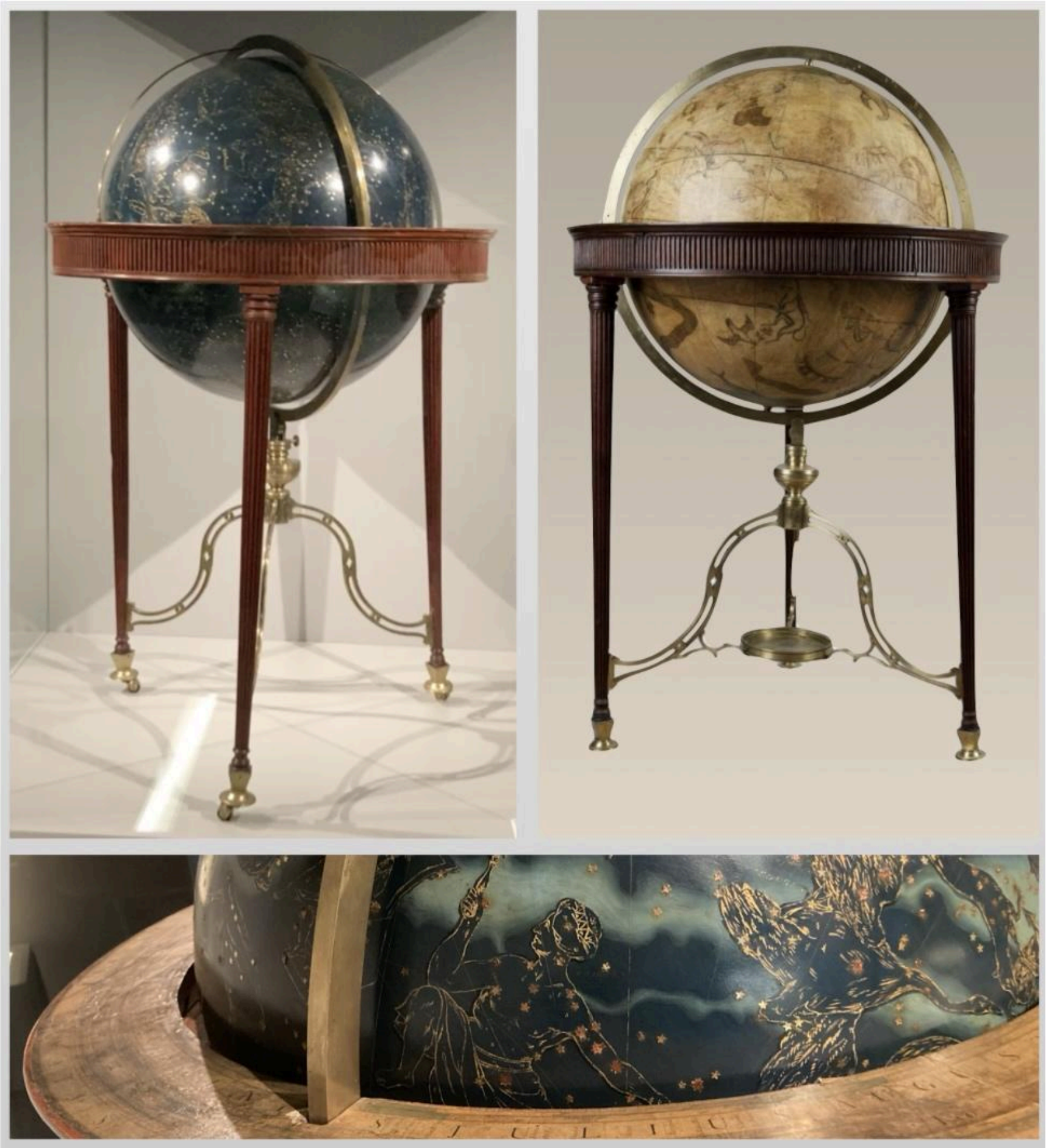

**Fig. 7**: Blue celestial globe from the Spanish National Library (top-left, detail below) and celestial globe by Dudley Adams at Museo de América in Madrid (top-right).

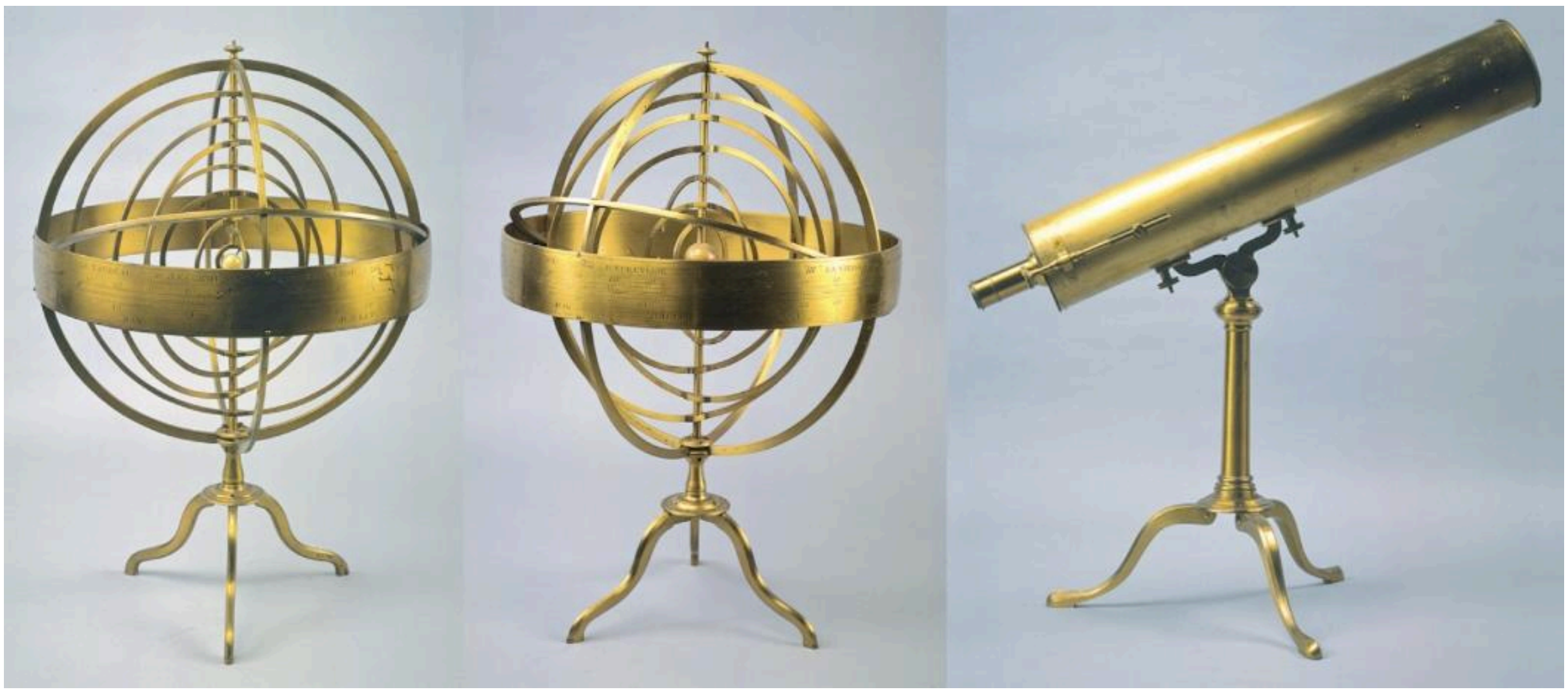

**Fig. 8**: Geocentric and heliocentric armillary spheres attributed to the workshop of instruments of the Royal Observatory of Madrid, next to a Gregorian telescope produced by Carlos Rodríguez and Mario Fernández in London in 1790.

## Other outstanding pieces from the catalog

The catalog brings to light other interesting globes. Regarding those produced in Spain, in addition to the globe by Tomás López (1770-80), there is a significant number of globes by Antonio Monfort (four terrestrial globes, two celestial globes, and four armillary spheres) which were produced in Barcelona around 1825-31 (Fig. 9). The tradition of globe production in Barcelona continued with Faustino Paluzie, who manufactured a few pieces at the very end of the 19th century and gave rise to a large production at the beginning of the 20th century (outside the scope of the present work). We refer the interested reader to Ovelar[58] for more details about the globe production of Faustino Paluzie. Falk[59] studied how Benjamín Tena produced in Vilafranca (close to Valencia) an ingenious terrestrial globe with a miniature orrery inside around 1900, held at the Whipple Museum of the History of Science in Cambridge; our catalog shows that the town hall of Vilafranca now also holds one of these globes.[60] The fascination for orreries was also reflected in the work of Billeter, who built two spectacular astronomical clocks between 1857 and 1869. José P. Morales also produced in Madrid at least one terrestrial globe in 1883.

The Spanish public collections include some interesting orreries which are sometimes part of sophisticated clocks. For example, the Royal Palace in Madrid displays a stunning clock with an orrery inside a blue celestial globe supported on the shoulders of a naked male figure. Traditionally known as the "Atlas clock," the sculpture must in fact represent the half-god Hercules,[61] as he wears a lion's skin around his waist, thus representing the moment when Hercules temporarily replaced the titan Atlas in the duty of holding up the heavens (Fig. 10). We owe this piece to the Swiss clock-maker Abraham Louis Breguet, famous for developing the first tourbillon mechanism and founder of one of the most prestigious clock companies. The manuscript celestial globe is profusely decorated with golden drawings of the constellations, including those introduced by Nicolas Louis de Lacaille in the mid-18th century (such as Antlia, Pyxis, or Microscopium). The celestial globe contains an orrery which has the peculiarity that it features a mechanism to mimic the elliptical orbits of the planets in the solar system.

The orrery includes the planet Uranus, therefore it must have been produced necessarily after 1781 (when William Herschel discovered it). Colón de Carvajal[62] dated the piece to the first third of the 19th century, while Aranda Huete[63] suggests a production date around 1800. The celestial globe and orrery bear strong similarities with a piece signed "Lemaire 1788" and held at Château de Malmaison;[64] in particular, they have exactly the same mechanism.[65] If we follow Seguin,[66] it seems likely that this is the clock that Abraham Louis Breguet initially produced for Louis XVI, but by the time the clock was finished in the early 1790s, the king was no longer a possible client. In 1797, the clock was offered to the French Convention, who rejected it given its high price, and ended up finding a buyer in the Spanish king Charles IV (probably mediated by Agustín de Betancourt). Judging by the similarity with the signed model of Malmaison, Breguet had likely subcontracted the execution of the planetary to Lemaire.[67]

The Royal Palace in Madrid and the neighbouring Galería de las Colecciones Reales display some more outstanding pieces which combine clockwork mechanisms with celestial globes, such as the beautiful "Chronos" clock by François-Louis Godon, an allegory of time. A blue celestial globe with golden stars and dark constellation drawings stands on a marble sculpture of an old man, with a winged boy holding a torch hovering over the celestial globe, where the actual clock is located. The so-called "Las cuatro fachadas" clock (*The four façades*) dates back to the beginning of the 18th century, and it was produced by Thomas Hildeyard. This highly sophisticated product contains a large number of clocks and mechanisms which indicate the time for different meridians, moon phases, sunrise and sunset, the equation of time, and even the tides for four different locations. A celestial globe made of glass crowns the piece. It contains a representation of the earth and the sun, with some constellation drawings engraved on the glass.

The latter is not the only vitreous celestial globe in Spanish public collections, nor is it necessary to resort to the historical collections of the Spanish monarchy to find one. In the Museo Marítimo del Cantábrico, as well as in four historical schools,[68] we find impressive orreries from the second half of the 19th century, which can be driven by a manual spring, held inside a spherical celestial globe made of glass. As studied in detail by José María Gómez (private communication), the orrery from Santander was produced in Paris by Émile Bertaux; even though it does not bear his signature, it is extremely similar to a signed model at the Musée national de la Marine in Paris.[69] The remaining four pieces closely follow this same model, with identical dimensions (45 cm diameter), despite some differences in the details of the orrery mechanism, which brings them closer to the piece from Paris.[70] Despite its fragility, glass might be regarded as the ideal material to represent the crystalline celestial sphere envisioned by Aristotle and Ptolemy; vitreous celestial globes underwent a process of popularization in the 18th century which continued throughout the 19th century,[71] often associated with clockwork orreries such as the one by Émile Bertaux from which we have identified five examples in Spain.

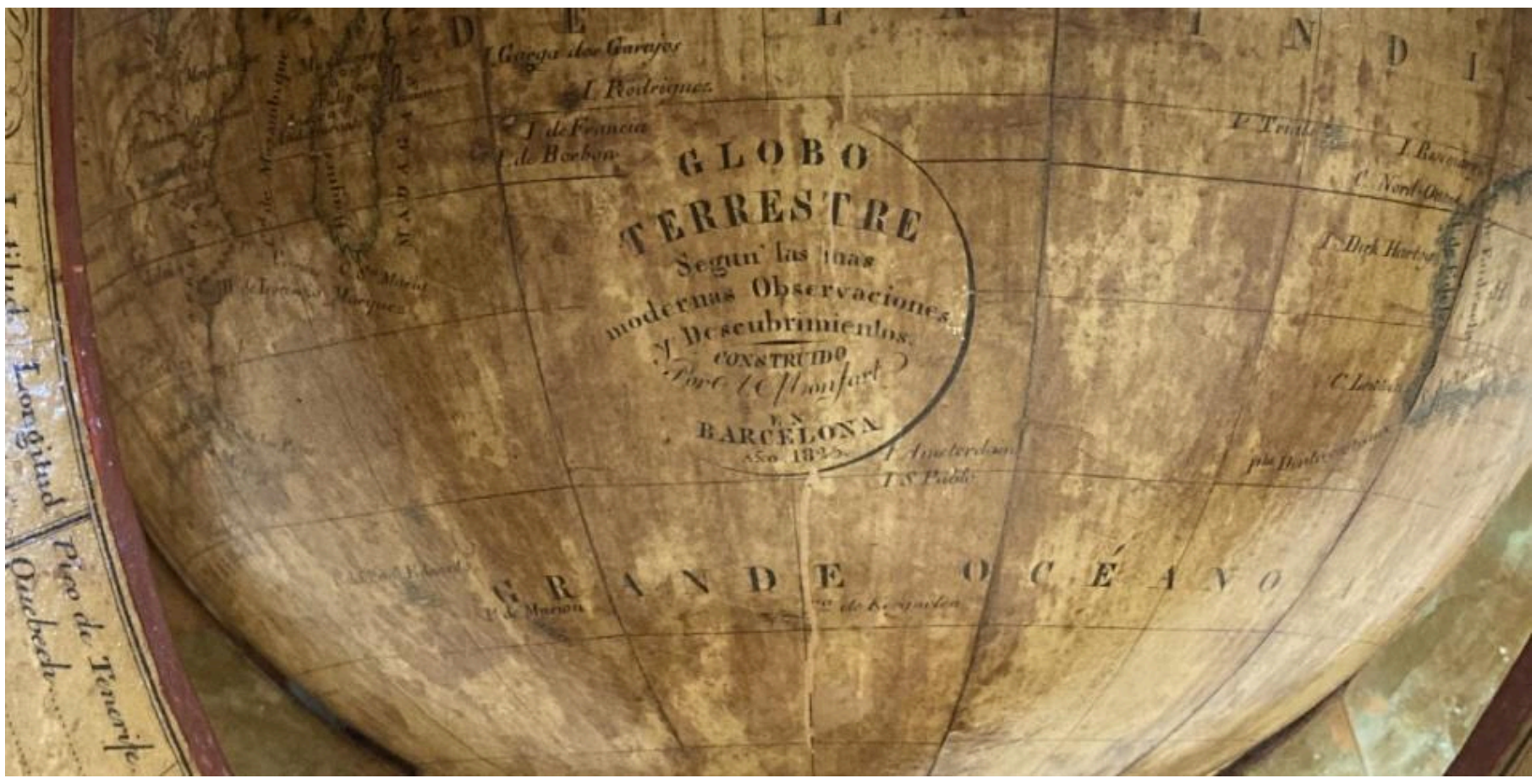

**Fig. 9**: Detail of the terrestrial globe by Antonio Monfort, produced in Barcelona in 1825 (Institut Cartogràfic i Geològic de Catalunya).

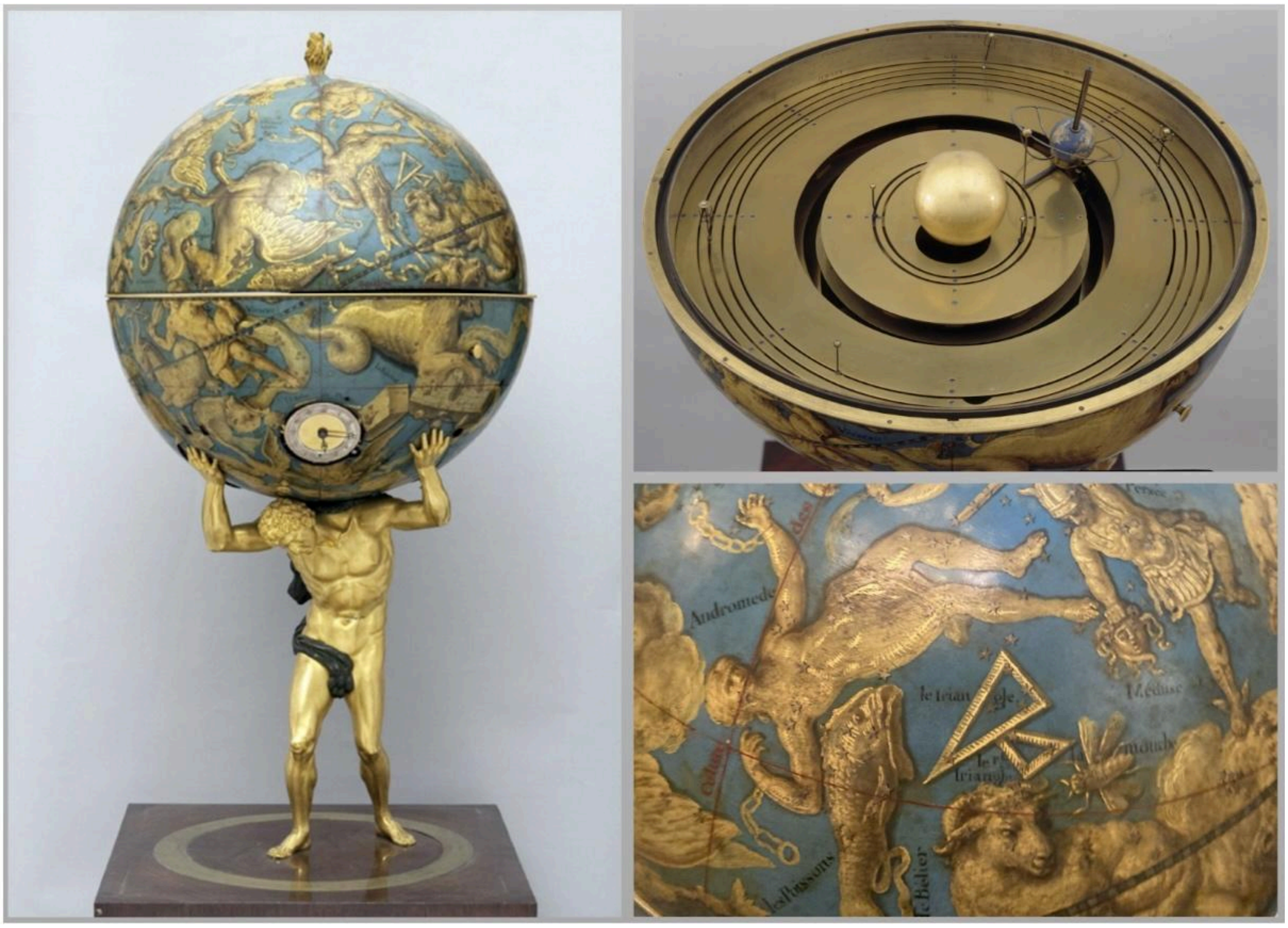

**Fig. 10**: General view and details of the so-called "Atlas clock" (which more likely represents Hercules) at the Royal Palace in Madrid.

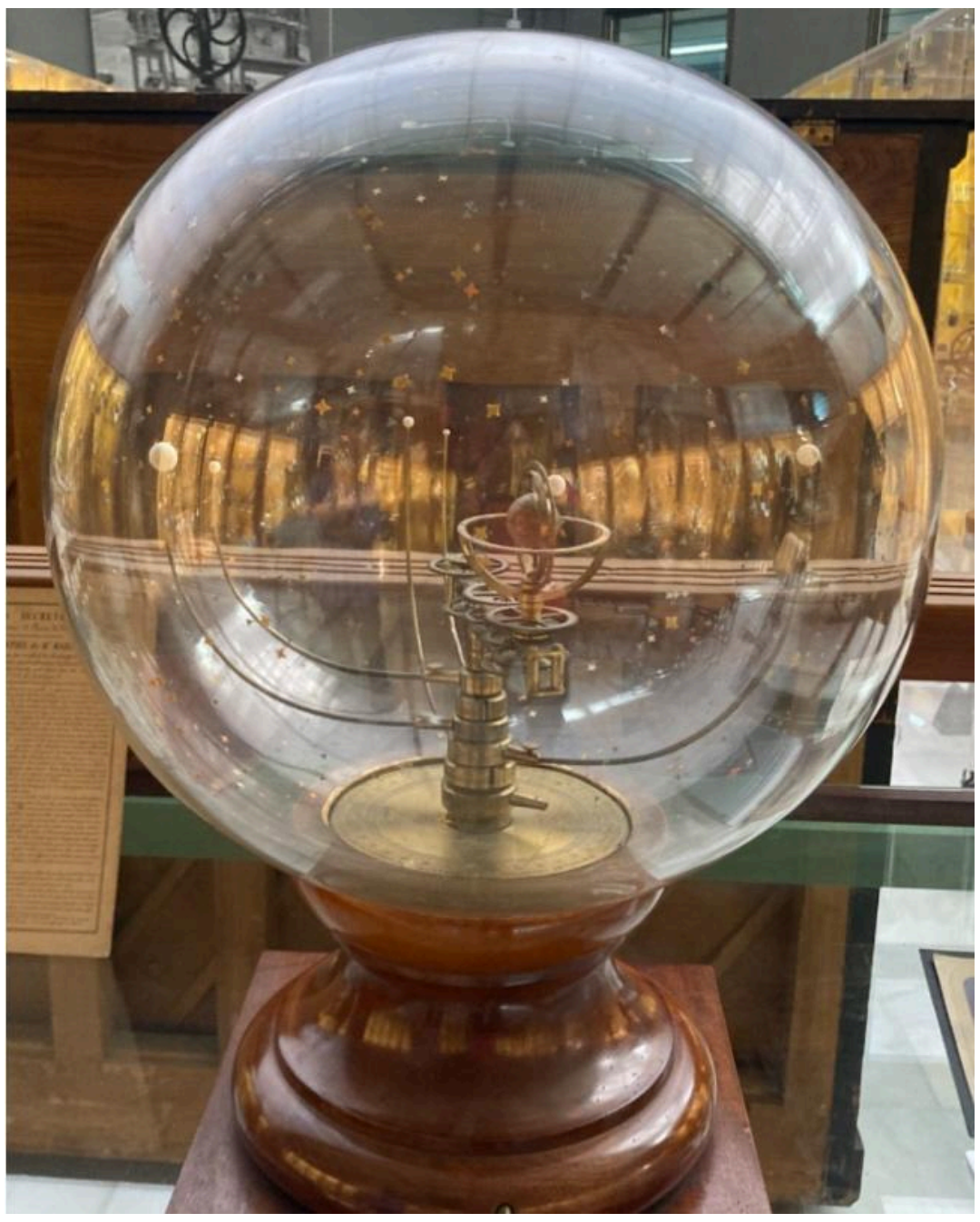

**Fig. 11**: Orrery by Émile Bertaux inside a celestial globe made of glass (IES Padre Suárez, Granada).

## Summary and conclusions

Despite its past political and maritime importance, Spain has never been regarded as a very relevant country in terms of globe production. There is written evidence of several globes produced in Spain at least since the Middle Ages, and a few Spanish globes from the 18th and 19th century have recently come to light. Yet, the relevance of Spain in terms of preserving globes produced not only in Spain but also elsewhere has not been examined in detail until now. This work tries to fill this gap by publishing the first catalog of historical globes currently held in Spanish public collections.

The catalog reveals quite a rich panorama in terms of variety of globes, country of origin, and epoch. As many as 199 pieces have been identified in Spanish public collections, including celestial and terrestrial globes, armillary spheres, planetaria, and orreries, as well as rare pieces such as tellurions and moon globes. We have identified examples of the best-known globe producers in the world, such as Mercator, Coronelli, or Blaeu, as well as families such as Adams, Cary, or Delamarche. Most globes have a French origin, followed by England, Spain, and the Netherlands. Celestial globes were highly popular through the 17th and 18th century, often appearing as celestial-terrestrial pairs, but their presence started to decline at the end of the 19th century in favour of individual terrestrial globes.

In this work, we have closely examined a few outstanding pieces. The metallic terrestrial globe from the Lázaro Galdiano Museum bears strong similarities with the “globe of Bailly” at the Morgan Library in New York (dated 1530), particularly regarding the cartography of America, southern Africa, and Asia. However, the departures are very significant in the cartographic details of Europe and northern Africa, and the inscriptions are plagued with misspellings. While it might be an early copy of the “globe of Bailly”, the fact that the inscription 1630 appears twice, instead of 1530, suggests that it might instead have been produced in the first half of the 17th century. We have also examined the oldest extant terrestrial globe produced in Spain, produced by Tomás López, for which we provide a narrower dating frame (1770-1780) than previously assumed, based on a compelling inscription found on the map near present-day Alaska. We have also put forward the hypothesis that the mysterious blue celestial globe from the Spanish National Library was produced in the decade of 1790 at the workshop of instruments of the Royal Observatory of Madrid, together with two armillary spheres that have traditionally been attributed to this workshop. For the blue celestial globe, whether it was produced at the workshop of the Observatory or not, it is clear that it is closely connected to a globe pair by Dudley Adams, which could have served as a model. Finally, we have commented on some interesting clockwork pieces, including manuscript celestial globes as well as vitreous globes. Some of them are associated with the historical collections of the Spanish monarchy, while others were acquired as educational tools at schools in the last third of the 19th century.

This is the first study that has compiled a catalog of globes, armillary spheres, and orreries preserved in Spanish public collections. This has allowed us to show that Spain preserves a significant number of spheres dating all the way back to the 16th century, and spanning a large number of types, sizes, and countries of origin. The catalog includes some particularly remarkable pieces which we have commented on. Thus, in conclusion, far from being a country lacking the production and conservation of globes, this study highlights Spain’s significance in this context.

**Acknowledgments:**

I would like to warmly thank Dr Thomas Horst (Bayerische Staatsbibliothek, Munich) for introducing me to the fascinating universe of globes and for encouraging me to pursue this project. I also gladly acknowledge helpful discussions with Catherine Hofmann (Bibliothèque nationale de France), Jan Mokre (Globenmuseum, Vienna), Carmen García Calatayud (Biblioteca Nacional de España), Luisa Martín Merás, Antonio Sánchez (Universidad Autónoma de Madrid), Inés Pérez Teresa (Universidad Complutense de Madrid), Amelia Aranda Huete (Patrimonio Nacional), César Ovelar, Agustín Hernando (Universidad de Barcelona), Pere Planesas, Sandra Sáenz-López (Universidad Autónoma de Madrid), José María Gómez, Paco Bellido, Miguel Ángel Muñecas, Chet van Duzer (University of Rochester), and Valerie Shrimplin. Furthermore, this project would not have been possible without the generous contribution from a wide range of people in charge of Spanish institutions who provided me with information and allowed me to inspect their collections. This includes museum curators, librarians, experts from cartographic departments, astronomers, university professors or heritage experts, as well as high-school teachers. I am extremely thankful to all of them: Miriam Font (Escuela de Estudios Árabes, Granada), Emilio Padilla (IES Padre Suárez, Granada), Purificación Marinetto (Museo de la Alhambra, Granada), Concha Mancebo (Universidad de Granada), Ricardo García Jurado (Universidad de Sevilla), Sara Lugo (Museo Pedagógico de Aragón), José Ramón García López (Museo Marítimo de Asturias), Liti García-Ramos (Universidad de La Laguna), José María Moreno (Museo Naval, Madrid), Teresa Juan (IES Cabrera Pinto, Santa Cruz de Tenerife), Lucía Fernández Granados, Raúl Gutiérrez Rodríguez (Museo Marítimo del Cantábrico), Sonsoles Soroa (IES Isabel de Castilla, Ávila), Margarita Becedas (Universidad de Salamanca), Inmaculada Sanz (IES Antonio Machado, Soria), Ernesto Vázquez Souza (Universidad de Valladolid), José Juan Morcillo (Museo del Niño, Albacete), José María Sánchez (Museo de Ciencias Castilla-La Mancha), Javier García Francisco (IES Brianda de Mendoza, Guadalajara), María José Sánchez-Cifuentes (IES El Greco, Toledo), David Domènech (Biblioteca Arús), Sílvia Ferrer, Concepció Isern (Biblioteca de Catalunya), Noelia Ramos (Institut Cartogràfic i Geològic de Catalunya) M.ª Dolors Jurado, Cristina Latorre (Museu Marítim de Barcelona), Anna Jané (Reial Acadèmia de Ciències i Arts de Barcelona), Inés Carballal (Museu Municipal de Nàutica del Masnou), Susana García (Castell de Peralada), Francesc Solé (Museu Nacional de la Ciència i la Tècnica de Catalunya), Mar Pérez Milla (Masia Cabanyes), Fernando J. Ballesteros (Universitat de València), Juan Manuel Cutrín (Universidade de Santiago de Compostela), Natalia Fraguas (Museo Provincial de Pontevedra), Yolanda Posedente (IES Sagasta, Logroño), Rosa María Martín Latorre, Ignacio de la Lastra (Museo Nacional de Ciencia y Tecnología), Virginia Albarrán, Almudena Pérez de Tudela, Javier Jordán de Urríes, Pilar Benito (Patrimonio Nacional), Mercedes Pasalodos (Biblioteca Nacional de España), Marcos Pavo, Antonio Basauri, Marta Montilla (Instituto Geográfico Nacional), María Ángeles Granados (Museo Arqueológico Nacional), María González Castañón (Museo de América), Sonia Fernández (Museo de Historia de Madrid), Carmen Riquelme (Museo del Aire y del Espacio), Carmen Espinosa (Museo Lázaro Galdiano), Sandra Rodríguez Bermejo (Congreso de los Diputados), Josefa Fuentes García (Senado), Asunción Miralles de Imperial (Real Academia de la Historia), Fabiola Azanza (Real Sociedad Económica Matritense de Amigos del País), Fuensanta González Montesinos, Beatriz González López-Plaza, Ángel Gómez Nicola (Universidad Complutense de Madrid), Andrés Nieto (IES Alfonso X el Sabio, Murcia), and Paco Conde (San Telmo Museoa, Donostia-San Sebastián).

**Notes on Contributor:**

**Miguel Querejeta** holds a degree in Physics from the Complutense University of Madrid, a degree in Spanish Language and Literature from UNED, and a PhD in Astrophysics from the University of Heidelberg. Since 2017, he has been a staff astronomer at the Royal Observatory of Madrid. His research focuses on nearby galaxies, and he has also carried out multiple projects on the history of astronomy and the intersections between art and astronomy. He has received several awards for his scientific outreach work.

**Table 2**: Catalog of globes in Spanish public collections (Type: "C" for celestial globes, "T" for terrestrial globes, "AS" for armillary spheres, "Orr" for orreries, "P" for planetaria, "Tell" for tellurions, "*" for clockwork pieces, as explained in the text).

| Region | Location | Institution | Type | Producer | Origin | Dia-meter | Date | Inventory number | Comment |
|---|---|---|---|---|---|---|---|---|---|
| Andalucía | Baeza | IES Antonio Machado | AS | Maison Delamarche? | France (Paris)? | 21 cm | ca. 1850-1900 | N/A | |
| Andalucía | Baeza | IES Antonio Machado | T | George Thomas | France (Paris) | 30 cm | ca. 1900 | N/A | |
| Andalucía | Cádiz | Universidad de Cádiz | T | Faustino Paluzie | Spain (Barcelona) | 25 cm | ca. 1880 | N/A | Publ.[72] |
| Andalucía | Cádiz | Universidad de Cádiz | AS | Faustino Paluzie | Spain (Barcelona) | 25 cm | ca. 1880 | N/A | Publ.[73] |
| Andalucía | Granada | Escuela de Estudios Árabes de Granada (CSIC) | T | Maison Delamarche | France (Paris) | 63 cm | 1870 | N/A | |
| Andalucía | Granada | IES Padre Suárez | Orr+C | Émile Bertaux | France (Paris) | 45 cm | ca. 1860 | N/A | |
| Andalucía | Granada | Museo de la Alhambra | C | Anonymous | Iran | 20 cm | 1700-1900? | Inv. R3971 | Bronze Islamic manuscript globe |
| Andalucía | Granada | Universidad de Granada (Hospital Real) | C | Alexandre Delamarche | France (Paris) | 63 cm | ca. 1875 | Inv. 1016 | Publ.[74] |
| Andalucía | San Fernando | Real Observatorio de la Armada | T | John & William Cary | England (London) | 53.5 cm | 1825 | ROA: 0097/PH | Publ.[75] |
| Andalucía | San Fernando | Real Observatorio de la Armada | C | John & William Cary | England (London) | 53.5 cm | 1827 | ROA: 0096/PH | Publ.[76] |
| Andalucía | San Fernando | Real Observatorio de la Armada | Pl | Anonymous | Unknown | 30 cm | ca. 1875-1900 | ROA: 0207/PH | |
| Andalucía | Sevilla | Universidad de Sevilla (Rectorado) | C | John & William Cary | England (London) | 53.5 cm | 1799 | Inv: 2206 | Publ.[77] |
| Andalucía | Sevilla | Universidad de Sevilla (Rectorado) | T | John & William Cary | England (London) | 53.5 cm | 1840 | Inv: 2207 | Publ.[78] |
| Aragón | Huesca | Museo Pedagógico de Aragón | AS | Anonymous | Unknown | 32 cm | ca. 1875-1900 | Inv: 00432 | Geocentric |
| Aragón | Huesca | Museo Pedagógico de Aragón | AS | Faustino Paluzie | Spain (Barcelona) | 31 cm | ca. 1875-1900 | Inv: 00270 | Geocentric. Publ.[79] |
| Aragón | Teruel | IES Vega del Turia | Tell | Anonymous | Unknown | 39 x 35 cm | ca. 1875-1900 | Inv. VT0250 | |
| Aragón | Teruel | Museo del Maestrazgo (Escuela de La Cuba) | AS | Faustino Paluzie | Spain (Barcelona) | 31 cm | ca. 1875-1900 | Inv: ESC. CUB. 0019 | Geocentric. Publ.[80] |
| Asturias | Luanco | Museo Marítimo de Asturias | AS | Faustino Paluzie | Spain (Barcelona) | 31 cm | ca. 1875-1900 | Inv. 105 | Geocentric. Publ.[81] |
| Canarias | La Laguna | Universidad de La Laguna (Biblioteca General y de Humanidades) | C | John Senex | England (London) | 68 cm | 1740 | N/A | Publ.[82] |
| Canarias | La Laguna | Universidad de La Laguna (Biblioteca General y de Humanidades) | T | John Senex | England (London) | 68 cm | 1740 | N/A | Publ.[83] |
| Canarias | La Laguna | Universidad de La Laguna (Facultad de Educación) | T | John & William Cary | England (London) | 48,5 cm | ca. 1800-1825 | N/A | Publ.[84] |
| Canarias | Santa Cruz de Tenerife | IES Cabrera Pinto | Tell | Léon Girod | France | 65 cm | ca. 1890 | Inv: 30 | Known as "cosmographe Girod" |
| Canarias | Santa Cruz de Tenerife | IES Cabrera Pinto | AS | Anonymous | France (Paris)? | 30cm | ca. 1840 | Inv: 80 | |
| Canarias | Santa Cruz de Tenerife | IES Cabrera Pinto | AS | Anonymous | Unknown | <41 cm | ca. 1850 | Inv: 55 | |
| Canarias | Santa Cruz de Tenerife | Real Sociedad Económica de Amigos del País de Tenerife | C | John & William Cary | England (London) | 45 cm | 1816 (corregidos en 1818) | N/A | Publ.[85] |
| Canarias | Santa Cruz de Tenerife | Real Sociedad Económica de Amigos del País de Tenerife | T | John & William Cary | England (London) | 45 cm | 1816 (corregidos en 1818) | N/A | Publ.[86] |
| Cantabria | Santander | Museo Marítimo del Cantábrico | T | George Adams | England (London) | 60 cm | 1789 | MMC CE005098 | Publ.[87] |
| Cantabria | Santander | Museo Marítimo del Cantábrico | C | Félix Delamarche | France (Paris) | 72 cm | ca. 1850 | MMC CE005632 | |
| Cantabria | Santander | Museo Marítimo del Cantábrico | T | Félix Delamarche | France (Paris) | 72 cm | 1870 | MMC CE005492 | |
| Cantabria | Santander | Museo Marítimo del Cantábrico | Orr+C | Émile Bertaux | France (Paris) | 45 cm | 1850-1900 | MMC CE004629 | |
| Castilla y León | Ávila | IES Isabel de Castilla | Orr+C | Émile Bertaux | France (Paris) | 45 cm | 1850-1900 | N/A | |
| Castilla y León | Salamanca | Universidad de Salamanca (Biblioteca Histórica) | C | Willem Janszoon Blaeu | Netherlands (Amsterdam) | 68 cm | ca. 1630-1650 | N/A | BLA V, state 3. Publ.[88] |
| Castilla y León | Salamanca | Universidad de Salamanca (Biblioteca Histórica) | T | Willem Janszoon Blaeu | Netherlands (Amsterdam) | 68 cm | ca. 1645-1648 | N/A | BLA V, state 4. Publ.[89] |
| Castilla y León | Salamanca | Universidad de Salamanca (Biblioteca Histórica) | C | Didier Robert de Vaugondy | France (Paris) | 45.5 cm | 1751 | N/A | Publ.[90] |
| Castilla y León | Salamanca | Universidad de Salamanca (Biblioteca Histórica) | T | L. C. Desnos; J. B. Nolin | France (Paris) | 44 cm | 1754 | N/A | |
| Castilla y León | Salamanca | Universidad de Salamanca (Biblioteca Histórica) | T | John Senex; Benjamin Hardon | England (London) | 66 cm | 1757 | N/A | Publ.[91] |
| Castilla y León | Salamanca | Universidad de Salamanca (Biblioteca Histórica) | C | Dudley Adams | England (London) | 42 cm | 1770-1800 | N/A | Publ.[92] |
| Castilla y León | Salamanca | Universidad de Salamanca (Biblioteca Histórica) | T | Antonio Monfort | Spain (Barcelona) | 23 cm | 1825 | N/A | Publ.[93] |
| Castilla y León | Salamanca | Universidad de Salamanca (Biblioteca Histórica) | AS | Antonio Monfort | Spain (Barcelona) | 23 cm | ca. 1820-1835 | N/A | Publ.[94] |
| Castilla y León | Salamanca | Universidad de Salamanca (Biblioteca Histórica) | C | Antonio Monfort | Spain (Barcelona) | 23 cm | 1830 | N/A | Publ.[95] |
| Castilla y León | Salamanca | Universidad de Salamanca (Biblioteca Histórica) | AS | Antonio Monfort | Spain (Barcelona) | 28 cm | 1831 | N/A | Publ.[96] |
| Castilla y León | Salamanca | Universidad de Salamanca (Facultad de Geografía e Historia) | T | Karl Christian Ludwig Adami, D. Reimer | Germany (Berlin) | 71 cm | 1871 | N/A | |
| Castilla y León | Soria | IES Antonio Machado | T | Lorrain / Bastien Ainé / Lapie | France (Paris) | 21.5 cm | ca. 1830-1840 | N/A | |
| Castilla y León | Valladolid | Universidad de Valladolid (Biblioteca General histórica de Santa Cruz) | C | Gerard Valk | Netherlands (Amsterdam) | 31 cm | 1700 | N/A | VAL II, state 1. Publ.[97] |
| Castilla y León | Valladolid | Universidad de Valladolid (Biblioteca General histórica de Santa Cruz) | T | Gerard & Leonard Valk | Netherlands (Amsterdam) | 31 cm | 1700 | N/A | VAL II, state 2. Publ.[98] |
| Castilla y León | Valladolid | Universidad de Valladolid (Biblioteca General histórica de Santa Cruz) | C | Didier Robert de Vaugondy | France (Paris) | 45.5 cm | 1784? | N/A | Publ.[99] |
| Castilla y León | Valladolid | Universidad de Valladolid (Biblioteca General histórica de Santa Cruz) | T | Didier Robert de Vaugondy | France (Paris) | 45.5 cm | 1784 | N/A | Publ.[100] |
| Castilla-La Mancha | Albacete | Museo del Niño "Juan Peralta" | C | Maison Delamarche | France (Paris) | 25,5 cm | 1870 | Inv. 776E | |
| Castilla-La Mancha | Albacete | Museo del Niño "Juan Peralta" | T | Faustino Paluzie | Spain (Barcelona) | 20 cm | 1900 | Inv. 1579E | Publ.[101] |
| Castilla-La Mancha | Cuenca | IES Histórico Alfonso VIII | AS | Anonymous | Unknown | 40 cm | ca. 1860 | Inv. MAP/GLB/7 | |
| Castilla-La Mancha | Cuenca | IES Histórico Alfonso VIII | T | R. Barbot | France (Paris) | 29,5 cm | ca. 1880 | Inv. MAP/GLB/2 | |
| Castilla-La Mancha | Cuenca | Museo de Ciencias Castilla-La Mancha | Orr+C | Émile Bertaux | France (Paris) | 45 cm | 1850-1900 | N/A | On loan from IES Histórico Alfonso VIII |
| Castilla-La Mancha | Guadalajara | IES Brianda de Mendoza | C | Anonymous (Delamarche?) | France (Paris)? | 24,5 cm | ca. 1850 | N/A | |
| Castilla-La Mancha | Toledo | IES Histórico El Greco | T | Charles-François Delamarche | France (Paris) | 32.6 cm | 1785 | Inv: 2378 | Publ.[102] |
| Castilla-La Mancha | Toledo | IES Histórico El Greco | C | Charles-François Delamarche | France (Paris) | 32.5 cm | 1805 | Inv: 2379 | Publ.[103] |
| Castilla-La Mancha | Toledo | IES Histórico El Greco | C | Ernst Schotte & Co. | Germany (Berlin) | 30 cm | 1875-1900 | Inv: 2591 | Publ.[104] |
| Cataluña | Barcelona | Biblioteca Arús | C | Émile Bertaux | France (Paris) | 33 cm | ca. 1875-1895 | N/A | |
| Cataluña | Barcelona | Biblioteca Arús | T | Maison Delamarche / Émile Bertaux | France (Paris) | 33 cm | ca. 1875-1895 | N/A | |
| Cataluña | Barcelona | Biblioteca de Catalunya | T | Antonio Monfort | Spain (Barcelona) | 23 cm | 1825 | N/A | Publ.[105] |
| Cataluña | Barcelona | Biblioteca de Catalunya | C | John Cary | England (London) | 53.5 cm | ca. 1831 | Inv.: R Globus 2 | Publ.[106] |
| Cataluña | Barcelona | Biblioteca de Catalunya | T | John Cary | England (London) | 53.5 cm | 1831 | Inv.: R Globus 1 | Publ.[107] |
| Cataluña | Barcelona | Institut Cartogràfic i Geològic de Catalunya | T | Antonio Monfort | Spain (Barcelona) | 23 cm | 1825 | RM.290293 | Publ.[108] |
| Cataluña | Barcelona | Institut Cartogràfic i Geològic de Catalunya | T | Carl Adami; Heinrich Kiepert; D. Reimer | Germany (Berlin) | 34 cm | 1874 | RM.235677 | Publ.[109] |
| Cataluña | Barcelona | Museu Marítim de Barcelona | C | Gerard & Leonard Valk | Netherlands (Amsterdam) | 39 cm | 1750? | Inv. 21154 | VAL III, state 4?, Publ.[110] |
| Cataluña | Barcelona | Museu Marítim de Barcelona | T | Charles-François Delamarche | France (Paris) | 32,6 cm | 1785 | Inv. 168 | Publ.[111] |
| Cataluña | Barcelona | Museu Marítim de Barcelona | AS | Charles-François Delamarche | France (Paris) | 46,5 cm | ca. 1800 | Inv. 169 | |
| Cataluña | Barcelona | Museu Marítim de Barcelona | AS | Charles-François Delamarche | France (Paris) | 32 cm | 1805 | Inv. 170 | Publ.[112] |
| Cataluña | Barcelona | Museu Marítim de Barcelona | C | Charles-François Delamarche | France (Paris) | 32,6 cm | 1805 | Inv. 167 | |
| Cataluña | Barcelona | Museu Marítim de Barcelona | C | Delamarche (Félix?) | France (Paris) | 25 cm | 1851 | Inv. 771 | |
| Cataluña | Barcelona | Museu Marítim de Barcelona | C | Alexandre Delamarche | France (Paris) | 63 cm | ca. 1875 | Inv. 5236 | |
| Cataluña | Barcelona | Reial Acadèmia de Ciències i Arts de Barcelona | Orr* | Alberto Billeter | Spain (Barcelona) | 60 cm | 1869 | N/A | Part of a large astronomical clock (total dimensions: 3.22 × 2.07 m) |
| Cataluña | El Masnou | Museu Municipal de Nàutica del Masnou | AS | Anonymous | Unknown | 32 cm | ca. 1899 | N/A | |
| Cataluña | Perelada | Biblioteca Castell de Peralada | AS | Anonymous | France (Paris) | 40 cm | 1800-1900 | Inv. 433 | Geocentric |
| Cataluña | Perelada | Biblioteca Castell de Peralada | AS | Anonymous | France (Paris) | 40 cm | 1800-1900 | Inv. 434 | Heliocentric |
| Cataluña | Terrassa | Museu Nacional de la Ciència i la Tècnica de Catalunya | AS | Antonio Monfort | Spain (Barcelona) | 28 cm? | 1829 | Inv. 17953 | Publ.[113] |
| Cataluña | Terrassa | Museu Nacional de la Ciència i la Tècnica de Catalunya | C | Antonio Monfort | Spain (Barcelona) | 23 cm | 1830 | Inv. 17954 | Publ.[114] |
| Cataluña | Vilanova i la Geltrú | Masia Cabanyes | T | Charles-François Delamarche | France (Paris) | 21 cm | 1799-1800 | Inv. 227 | |
| Cataluña | Vilanova i la Geltrú | Masia Cabanyes | AS | Charles-François Delamarche? | France (Paris) | 21 cm | ca. 1800? | Inv. 228 | |
| Comunidad Valenciana | València | Biblioteca Histórica, Universitat de València | C | Willem Janszoon Blaeu | Netherlands (Amsterdam) | 68 cm | ca. 1630-1650 | N/A | BLA V, state 3. Publ.[115] |
| Comunidad Valenciana | València | Biblioteca Histórica, Universitat de València | T | Willem Janszoon Blaeu | Netherlands (Amsterdam) | 68 cm | ca. 1645-1648 | N/A | BLA V, state 4. Publ.[116] |
| Comunidad Valenciana | València | Observatori Astronòmic, Universitat de València | AS | Antonio Monfort | Spain (Barcelona) | ca. 30 cm | 1831 | N/A | Heliocentric. Publ.[117] |
| Comunidad Valenciana | Vilafranca | Ajuntament de Vilafranca | T+Orr | Benjamín Tena | Spain (Villafranca) | 25 cm | ca. 1900 | N/A | Terrestrial globe with orrery inside. Publ.[118] |
| Galicia | Pontevedra | Museo Provincial Pontevedra | AS | Anonymous | Unknown | 40 cm | ca. 1800-1850 | Inv. 001778 | On loan from Instituto Nacional de Enseñanza Media de Pontevedra |
| Galicia | Bueu | Museo Massó | T+C | Nicholas Lane | England (London) | 8,5 cm | 1776 | N/A | Publ.[119] |
| Galicia | Bueu | Museo Massó | C | A.R. Fremin | France (Paris) | 26 cm | 1842 | CE28 | |
| Galicia | Bueu | Museo Massó | T | Charles Dien | France (Paris) | 25 cm | 1843 | CE19 | Publ.[120] |
| Galicia | Bueu | Museo Massó | T | Anonymous | Unknown | 45 cm | 1800-1900 | CE105 | |
| Galicia | Bueu | Museo Massó | Pl | Anonymous | France? | 30 cm | 1800-1900 | Inv.: CE10 | |
| Galicia | Bueu | Museo Massó | Pl | Anonymous | Spain? | 32 cm | ca. 1800-1900 | Inv.: CE14 | |
| Galicia | Bueu | Museo Massó | C | Anonymous | Unknown | 24 cm | ca. 1850-1890 | CE24 | |
| Galicia | Bueu | Museo Massó | C | Charles Dien | France (Paris) | 25 cm | 1874 | CE6 | |
| Galicia | Bueu | Museo Massó | AS | Faustino Paluzie | Spain (Barcelona) | 31 cm | ca. 1860-1890 | CE25 | Geocentric. Publ.[121] |
| Galicia | Santiago de Compostela | Museo Pedagóxico de Galicia | Pl | Anonymous | France? | 29 cm | 1878 | N/A | On loan from IES Xelmírez I |
| Galicia | Santiago de Compostela | Museo Pedagóxico de Galicia | T | Anonymous | Unknown | 110 cm | ca. 1880 | N/A | |
| Galicia | Santiago de Compostela | Museo Pedagóxico de Galicia | AS | Mang | Germany | 29 cm | ca. 1880 | Inv.: 117 | |
| Galicia | Santiago de Compostela | Museo Pedagóxico de Galicia | AS | J. Forest | France (Paris) | 35 cm | ca. 1880 | Inv.: 118 | |
| Galicia | Santiago de Compostela | Museo Pedagóxico de Galicia | AS | J. Forest | France (Paris) | 28.5 cm | ca. 1880 | Inv.: 200 | |
| Galicia | Santiago de Compostela | Museo Pedagóxico de Galicia | AS | Anonymous | Unknown | 28.5 cm | ca. 1880 | Inv.: 1356 | |
| Galicia | Santiago de Compostela | Museo Pedagóxico de Galicia | C | Faustino Paluzie | Spain (Barcelona) | 27 cm | ca. 1890 | Inv.: 143 | Publ.[122] |
| Galicia | Santiago de Compostela | Universidade de Santiago de Compostela | C | Félix Delamarche | France (Paris) | 45 cm | 1818 | Inv: IBC0000357 | |
| Galicia | Santiago de Compostela | Universidade de Santiago de Compostela | T | Félix Delamarche | France (Paris) | 45 cm | 1818 | Inv: IBC0000358 | |
| La Rioja | Logroño | IES Sagasta | T | Charles Dien | France (Paris) | 25 cm | ca. 1840 | N/A | Publ.[123] |
| Madrid | Alcobendas | Museo Nacional de Ciencia y Tecnología | T+C | Joseph Moxon | England (London) | 7 cm | ca. 1670 | CE2005/034 /0002 | Publ.[124] |
| Madrid | Alcobendas | Museo Nacional de Ciencia y Tecnología | T | John Senex | England (London) | 68 cm | ca. 1720-1730 | CE2003/019 /0002 | Publ.[125] |
| Madrid | Alcobendas | Museo Nacional de Ciencia y Tecnología | C | John Senex | England (London) | 68 cm | ca. 1740? | CE2003/019/0001 | Publ.[126] |
| Madrid | Alcobendas | Museo Nacional de Ciencia y Tecnología | AS | George Adams | England (London) | 31 cm | 1734-1773 | CE1985/004/0835 | On loan from IES San Isidro. Publ.[127] |
| Madrid | Alcobendas | Museo Nacional de Ciencia y Tecnología | Tell | George Adams | England (London) | 37 cm | 1734-1773 | DO1995/029/0001 | Publ.[128] |
| Madrid | Alcobendas | Museo Nacional de Ciencia y Tecnología | C | George Adams | England (London) | 46 cm | 1760-1766 | CE2002/019/0001 | Publ.[129] |
| Madrid | Alcobendas | Museo Nacional de Ciencia y Tecnología | T+C | John Newton | England (London) | 7 cm | ca. 1783-1818 | CE2005/034/0003 | Publ.[130] |
| Madrid | Alcobendas | Museo Nacional de Ciencia y Tecnología | C | John & William Newton | England (London) | 38 cm | 1818? | CE2005/033/0001 | Publ.[131] |
| Madrid | Alcobendas | Museo Nacional de Ciencia y Tecnología | T | Antonio Monfort | Spain (Barcelona) | 23 cm | 1825 | CE1986/006/1219 | Publ.[132] |
| Madrid | Aranjuez | Palacio Real de Aranjuez | T* | Louis Charles Desnos | France (Paris) | 30 cm | 1782 | Inv.: 10012729 | Sophisticated clock containing a terrestrial globe. Publ.[133] |
| Madrid | Aranjuez | Palacio Real de Aranjuez | T | Carl Adami; Heinrich Kiepert; D. Reimer | Germany (Berlin) | 34 cm | 1870 | Inv. 10086660 | Publ.[134] |
| Madrid | Aranjuez | Real Casa del Labrador | Orr* | Zacharie Raingo | France (Paris) | 19 x 25 cm | ca. 1800-1810 | Inv.: 10003660 | Publ.[135] |
| Madrid | Madrid | Biblioteca Nacional de España | T | Tomás López | Spain (Madrid) | 30 cm | 1770-1780 | GM/Globo 1 | Publ.[136] |
| Madrid | Madrid | Biblioteca Nacional de España | C | Anonymous | England (London)? | 68 cm | ca. 1795 ? | BNEM CE0365 | Manuscript globe. Publ.[137] |
| Madrid | Madrid | Colegio Público San Ildefonso | Orr+C | Émile Bertaux | France (Paris) | 45 cm | 1850-1887 | N/A | |
| Madrid | Madrid | Galería de las Colecciones Reales | C* | Thomas Hildeyard | Belgium (Liège) | 13 cm | ca. 1725 | Inv.: 10055705 | Astronomical clock known as "Las cuatro fachadas". Publ.[138] |
| Madrid | Madrid | Instituto Geográfico Nacional | C | Dudley Adams | England (London) | 25 cm | 1804 | Inv.: GLOBO008 | Publ.[139] |
| Madrid | Madrid | Instituto Geográfico Nacional | T | John & William Cary | England (London) | 30.5 cm | 1820 | Inv.: GLOBO009 | Publ.[140] |
| Madrid | Madrid | Instituto Geográfico Nacional | T | George & John Cary | England (London) | 30.5 cm | 1828 | N/A | Publ.[141] |
| Madrid | Madrid | Instituto Geográfico Nacional | T | Josiah Loring | USA (Boston) | 30.5 cm | 1833 | Inv.: GLOBO006 | Publ.[142] |
| Madrid | Madrid | Instituto Geográfico Nacional | C | Josiah Loring | USA (Boston) | 30.5 cm | 1835 | Inv.: GLOBO007 | Publ.[143] |
| Madrid | Madrid | Instituto Geográfico Nacional | T | Samuel S. Edkins; J. Souter | England (London) | 30.5 cm | 1837 | Inv.: GLOBO012 | Publ.[144] |
| Madrid | Madrid | Instituto Geográfico Nacional | T | Ernst Schotte | Germany (Berlin) | 39 cm | ca. 1880 | N/A | |
| Madrid | Madrid | Instituto Geográfico Nacional | T | D. Windels | Belgium (Bruxelles) | 32 cm | ca. 1890 | Inv.: GLOBO013 | |
| Madrid | Madrid | Instituto Geográfico Nacional | T | J. Lebègue | Belgium (Bruxelles) | 33 cm | ca. 1890 | Inv.: GLOBO014 | Publ.[145] |
| Madrid | Madrid | Instituto Geográfico Nacional | T | Ludwig J. Heymann; Henry Langue | Germany (Berlin) | 36 cm | ca. 1890 | Inv.: GLOBO015 | |
| Madrid | Madrid | Instituto Geográfico Nacional | T | George Thomas; Edmond Dubail | France (Paris) | 38 cm | ca. 1890 | Inv.: GLOBO010 | |
| Madrid | Madrid | Instituto Geográfico Nacional | C | Dietrich Reimer; Dr. H. Gewecke | Germany (Berlin) | 34 cm | ca. 1892 | Inv.: GLOBO011 | |
| Madrid | Madrid | Museo Arqueológico Nacional | C | Gerard Mercator | Belgium (Leuven) | 42 cm | 1551 | 2014/91/1 | Publ.[146] |
| Madrid | Madrid | Museo Arqueológico Nacional | T | Gerard Mercator | Belgium (Leuven) | 42 cm | 1541 | 2014/91/2 | Publ.[147] |
| Madrid | Madrid | Museo Arqueológico Nacional | C | Willem Janszoon Blaeu | Netherlands (Amsterdam) | 34 cm | ca. 1621-1638 | 1995/108/4 | BLA I, state 3. Publ.[148] |
| Madrid | Madrid | Museo Arqueológico Nacional | C | Willem Janszoon Blaeu | Netherlands (Amsterdam) | 68 cm | ca. 1630-1650 | Inv. 1995/108/2 | BLA V, state 3. Publ.[149] |
| Madrid | Madrid | Museo Arqueológico Nacional | T | Willem Janszoon Blaeu | Netherlands (Amsterdam) | 68 cm | ca. 1645-1648 | Inv. 1995/108/1 | BLA V, state 4. Publ.[150] |
| Madrid | Madrid | Museo Arqueológico Nacional | T | Guillaume Delisle | France (Paris) | 31 cm | 1700 | 1995/108/3 | Publ.[151] |
| Madrid | Madrid | Museo Arqueológico Nacional | C | Johan Gabriel Doppelmayr; Johann Georg Puschner | Germany (Nüremberg) | 20 cm | 1736 | Inv: 55970 | Publ.[152] |
| Madrid | Madrid | Museo de América | C | Dudley Adams | England (London) | 68 cm | ca. 1793 | Inv: 06656 | Publ.[153] |
| Madrid | Madrid | Museo de América | T | Dudley Adams | England (London) | 68 cm | ca. 1793 | Inv: 06657 | Publ.[154] |
| Madrid | Madrid | Museo de América | C | Thomas Malby | England (London) | 45 cm | ca. 1850 | Inv. 1981/08/1 | Publ.[155] |
| Madrid | Madrid | Museo de América | T | Thomas Malby | England (London) | 45 cm | 1876 | Inv. 1981/08/2 | Publ.[156] |
| Madrid | Madrid | Museo de Historia de Madrid | AS | Celedonio Rostriaga | Spain (Madrid) | 35 cm | 1798 | Inv.: 00001.768 | Publ.[157] |
| Madrid | Madrid | Museo del Aire y del Espacio | C | Dietrich Reimer | Germany (Berlin) | 34 cm | ca. 1852-1900 | MAA 857 | |
| Madrid | Madrid | Museo del Traje | T | Ernst Schotte | Germany (Berlin) | 33 cm | ca. 1900 | CE114221 | |
| Madrid | Madrid | Museo Lázaro Galdiano | T* | Robert de Bailly | Germany | 14 cm | 1530/1630? | Inv. 01481 | Part of a cup, including a manuscript terrestrial globe. Publ.[158] |
| Madrid | Madrid | Museo Nacional de Ciencia y Tecnología | C | Willem Janszoon Blaeu | Netherlands | 34 cm | ca. 1621-1638 | CE2002/032/0001 | BLA I, state 3. Publ.[159] |
| Madrid | Madrid | Museo Nacional de Ciencia y Tecnología | AS | Anonymous | Unknown | 9.7 cm | 1600-1800 | CE2009/032/0003 | |
| Madrid | Madrid | Museo Nacional de Ciencia y Tecnología | Orr | William Harris | England (London) | 35 cm | ca. 1816-1839 | CE1992/014/0002 | Publ.[160] |
| Madrid | Madrid | Museo Nacional de Ciencia y Tecnología | AS | Delamarche | France (Paris) | 21 cm | ca. 1830-1840 | CE1986/006/1220 | |
| Madrid | Madrid | Museo Naval | AS | Unknown | France (Paris)? | 22,2 cm | 1600-1700 | MNM-3502 | Geocentric (brass) |
| Madrid | Madrid | Museo Naval | T | Vincenzo Coronelli | France (Paris) | 108 cm | 1688 | I-548 | Publ.[161] |
| Madrid | Madrid | Museo Naval | C | Vincenzo Coronelli | France (Paris) | 108 cm | 1693 | I-1839 | Publ.[162] |
| Madrid | Madrid | Museo Naval | AS | Unknown | Unknown | 36,5 cm | 1600-1800 | MNM-117 | Heliocentric (brass) |
| Madrid | Madrid | Museo Naval | AS | Unknown | France (Paris)? | 37 cm | 1700-1800 | MNM-785 | Geocentric |
| Madrid | Madrid | Museo Naval | T | Matthaeus Seutter | Germany (Augusburg) | 20,5 cm | 1757 | Inv. 1105 | Publ.[163] |
| Madrid | Madrid | Museo Naval | T | John & William Cary | England (London) | 7,6 cm | 1791 | Inv. 1712 | Publ.[164] |
| Madrid | Madrid | Museo Naval | T | Delamarche | France (Paris) | 32 cm | 1804 | I-2375 | Publ.[165] |
| Madrid | Madrid | Museo Naval | C | A. R. Fremin | France (Paris) | 28 cm | 1842 | I-1573 | |
| Madrid | Madrid | Museo Naval | C | Delamarche | France (Paris) | 26 cm | 1865? | I-1841 | |
| Madrid | Madrid | Museo Naval | T | Abel / Klinger | Germany (Nuremberg) | 7,5 cm | 1800-1900 | I-1314 | |
| Madrid | Madrid | Museo Naval | T | Delamarche / Chartier | France (Paris) | 33,5 cm | 1865 | Inv. 784 | |
| Madrid | Madrid | Museo Naval | T | Anonymous | Unknown | 7,2 cm | 1900? | I-1871 | |
| Madrid | Madrid | Museo Sorolla | T | Ernst Schotte & Co. | Germany (Berlin) | 40 cm | ca. 1875-1900 | Inv.: 90081 | |
| Madrid | Madrid | Palacio de la Zarzuela | T* | Jean-Simon Bourdier | France (Paris) | 18 cm | ca. 1800 | Inv.: 10055806 | Sophisticated clock with a terrestrial globe |
| Madrid | Madrid | Palacio del Congreso de los Diputados | Orr* | Alberto Billeter | Spain (Barcelona) | 60 cm | 1857 | N/A | Part of a large astronomical clock |
| Madrid | Madrid | Palacio del Senado | T | George Adams | England (London) | 46 cm | ca. 1766 | N/A | Publ.[166] |
| Madrid | Madrid | Palacio Real de Madrid | C | Jean-Baptiste Fortin | France (Paris) | 22 cm | 1780 | Inv. 10086658 | |
| Madrid | Madrid | Palacio Real de Madrid | C+Orr * | Abraham-Louis Breguet | Switzerland (Neuchâtel) | 50 cm | ca. 1790-1795 | Inv.: 10003300 | "Atlas" clock. Publ.[167] |
| Madrid | Madrid | Palacio Real de Madrid | C* | Francisco Luis Godon | France (Paris) | 29 cm | ca. 1798-1800 | Inv.: 10002970 | "The Time" or "Cupid" clock. Publ.[168] |
| Madrid | Madrid | Palacio Real de Madrid | T* | Anonymous | Unknown | 7 cm | ca. 1800 | Inv.: 10003257 | Clock known as "Ceres". Publ.[169] |
| Madrid | Madrid | Palacio Real de Madrid | T | Félix Delamarche | France (Paris) | 25 cm | 1821 | Inv. 10086659 | |
| Madrid | Madrid | Palacio Real de Madrid | T | Maison Delamarche | France (Paris) | 33 cm | 1861 | Inv. 10086661 | |
| Madrid | Madrid | Palacio Real de Madrid | T | Charles Larochette; Louis Bonnefont | France (Paris) | 50 cm | ca. 1870 | Inv. 10086662 | |
| Madrid | Madrid | Real Academia de la Historia | C | Willem Janszoon Blaeu | Netherlands | 34 cm | ca. 1621-1638 | N.º reg.: 1104 | BLA I, state 3. Publ.[170] |
| Madrid | Madrid | Real Academia de la Historia | T | Willem Janszoon Blaeu; Johannes van Ceulen | Netherlands | 34 cm | 1682 | N.º reg.: 1105 | BLA I, state 4. Publ.[171] |
| Madrid | Madrid | Real Academia de la Historia | T | Jean-Baptiste Fortin | France (Paris) | 34 cm | 1768 | N.º reg.: 1106 | |
| Madrid | Madrid | Real Academia de la Historia | T | Instituto Geografico Guido Cora | Torino | 100 cm | 1883 | N.º reg.: 1107 | |
| Madrid | Madrid | Real Observatorio de Madrid | Moon | John Russell | England (London) | 32,5 cm | ca. 1797 | Inv.: 87/1/04 | Selenographia (lunar sphere). Publ.[172] |
| Madrid | Madrid | Real Observatorio de Madrid | AS | Anonymous | Spain (Madrid)? | 39,5 cm | ca. 1800 | Inv.: 87/1/02 | Geocentric (brass). Publ.[173] |
| Madrid | Madrid | Real Observatorio de Madrid | AS | Anonymous | Spain (Madrid)? | 39,5 cm | ca. 1800 | Inv.: 87/1/03 | Helioentric (brass). Publ.[174] |
| Madrid | Madrid | Real Sociedad Económica Matritense de Amigos del País | AS | Diego Rostriaga | Spain (Madrid) | 31 cm | 1764-1783 | N/A | Geocentric (brass) |
| Madrid | Madrid | Real Sociedad Económica Matritense de Amigos del País | AS | Celedonio Rostriaga | Spain (Madrid) | 39 cm | 1794 | N/A | Helioentric (brass) |
| Madrid | Madrid | Universidad Complutense de Madrid (Facultad de Bellas Artes) | C | Didier Robert de Vaugondy | France (Paris) | 45,5 cm | 1751 | Inv.: CUC000571 | |
| Madrid | Madrid | Universidad Complutense de Madrid (Facultad de Bellas Artes) | T | Didier Robert de Vaugondy | France (Paris) | 45,5 cm | 1751 | Inv.: CUC000591 | |
| Madrid | Madrid | Universidad Complutense de Madrid (Facultad de Física) | T | José P. Morales | Spain (Madrid) | 69,5 cm | 1883 | Inv.: UCM 410782 | Publ.[175] |
| Madrid | Madrid | Universidad Complutense de Madrid (Museo de Astronomía y Geodesia) | Tell | A. H. Dufour | France (Paris) | 39 x 40 cm | 1831 | N/A | A. H. Dufour signed on the terrestrial globe |
| Madrid | Madrid | Universidad Complutense de Madrid (Museo de Astronomía y Geodesia) | AS | Maison Delamarche? | France (Paris) | 23 cm | 1850? | N/A | Geocentric |
| Madrid | Madrid | Universidad Complutense de Madrid (Museo de Astronomía y Geodesia) | Orr | Anonymous | France (Paris) | 33 x 57 cm | 1856 | N/A | |
| Madrid | Madrid | Universidad Complutense de Madrid (Museo de Astronomía y Geodesia) | C | M. Ch. Simon; Gustav Bagge; Louis Wuhrer | France (Paris) | 25,5 cm | 1880 | N/A | |
| Madrid | San Lorenzo de El Escorial | Casita del Infante | T* | Anonymous | France (Paris) | 18 cm | 1775-1800 | Inv. 10033621 | Sophisticated clock containing a terrestrial globe |
| Madrid | San Lorenzo de El Escorial | Real Monasterio de San Lorenzo de El Escorial | AS | Antonio Santucci | Italy (Florence) | 100 cm | ca. 1582 | Inv. 10034500 | Publ.[176] |
| Madrid | San Lorenzo de El Escorial | Real Monasterio de San Lorenzo de El Escorial | AS | Michel Coignet; Ferdinand Arsenius | Belgium (Antwerp) | 36 cm | ca. 1590 | Inv. 10014301 | Geocentric (brass) |
| Madrid | San Lorenzo de El Escorial | Real Monasterio de San Lorenzo de El Escorial | C | Willem Janszoon Blaeu | Netherlands (Amsterdam) | 68 cm | ca. 1630-1650 | Inv. 10034541 | BLA V, state 3. Publ.[177] |
| Madrid | San Lorenzo de El Escorial | Real Monasterio de San Lorenzo de El Escorial | T | Willem Janszoon Blaeu | Netherlands (Amsterdam) | 68 cm | ca. 1645-1648 | Inv. 10034544 | BLA V, state 4. Publ.[178] |
| Madrid | San Lorenzo de El Escorial | Real Monasterio de San Lorenzo de El Escorial | C | Charles Price | England (London) | 40 cm | 1716 | Inv. 10034542 | Publ.[179] |
| Madrid | San Lorenzo de El Escorial | Real Monasterio de San Lorenzo de El Escorial | T | Charles Price | England (London) | 40 cm | ca. 1716 | Inv. 10034543 | Publ.[180] |
| Madrid | San Lorenzo de El Escorial | Real Monasterio de San Lorenzo de El Escorial | C* | Anonymous | Unknown | 6 cm | ca. 1800-1830 | Cat. 246 | Clock known as "Orfeo" containing a celestial globe |
| Murcia | Murcia | IES Alfonso X el Sabio | Tell | Léon Girod | France | 65 cm | 1887 | N/A | Known as "cosmographe Girod" |
| País Vasco | Donostia-San Sebastián | San Telmo Museoa | T | J. Lebègue | Belgium (Bruxelles) | 60 cm | ca. 1880 | N/A | Publ.[181] |

## Notes:

1. J. Mokre and P.E. Allmayer-Beck (eds.), *Rund um den Globus: Erd- und Himmelsgloben und ihre Darstellungen* (Vienna: Bibliophile Edition, 2008); C. Hofmann and F. Nawrocki (eds.), *Le monde en sphères* (Paris: BnF éditions, 2019).
2. I. Ridpath, *Star Tales* (Cambridge: Lutterworth Press, 2018).
3. Hofmann and Nawrocki, *op. cit.* (Note 1), pp. 52–79.
4. S. Sumira, *The Art and History of Globes* (London: The British Library Publishing, 2014).
5. P. Van der Krogt, *Globi Neerlandici. The Production of Globes in the Low Countries* (Utrecht: Hes & de Graaf, 1993).
6. J.R. Millburn, *Adams of Fleet Street, Instrument Makers to King George III* (London: Routledge, 2000).
7. P.E. Allmayer-Beck (ed.), *Modelle der Welt: Erd- und Himmelsgloben* (Vienna: Christian Brandstätter Verlag, 1997).
8. Hofmann and F. Nawrocki, *op. cit.* (Note 1), pp. 152–9.
9. E.L. Stevenson, *Terrestrial and Celestial Globes*, (New Haven: Yale University Press, 1921).
10. Austria: Allmayer-Beck, *op. cit.* (Note 7); Bavaria: A. Fauser, *Ältere Erd- und Himmelsgloben in Bayern* (Stuttgart: Schuler Verlagsgesellschaft, 1964), A. Fauser, "Ältere Erd- und Himmelsgloben in Bayern (Nachträge zu dem 1964 erschienenen Buch)," *Der Globusfreund*, 31/32 (1983), 107–28; Denmark: R. Kejlbo, "Prov. Liste alter Globen in Dänemark," *Der Globusfreund*, 17 (1968), 22–7; France: G. Duprat, "Les globes terrestres et célestes en France," *Der Globusfreund*, 21/23 (1973), 198-225; Hessen: W. Kummer, "Liste alter Globen im Bundesland Hessen und aus einer Sammlung in Ingelheim in Rheinhessen," *Der Globusfreund* 28/29 (1980), 67–112, W. Kummer, "Liste alter Globen im Bundesland Hessen und aus einer Sammlung in Ingelheim in Rheinhessen: 2. Teil," *Der Globusfreund*, 31/32 (1983), 15–68; Hungary: Z. Ambrus-Fallenbüchl, "Geschichte und Liste der ungarischen Globen," *Der Globusfreund*, 21/23 (1973), 243–4; Japan: H. Kawamura et al., "List of Old Globes in Japan," *Der Globusfreund*, 38/39 (1990), 173–7; the Netherlands: P. van der Krogt, "List of Old Globes in the Netherlands," *Der Globusfreund*, 31/32 (1983), 78–106, P. van der Krogt, "List of Old Globes in the Netherlands: Additions to the list published in "Der Globusfreund 11 31-32," *Der Globusfreund*, 33/34 (1985), 143–7; Norway: T.E. Rössaak, "List of Old Globes in Norway," *Der Globusfreund*, 35/37 (1987), 255–8; Poland: B. Olszewicz, "Alte Globen in Polen," *Der Globusfreund*, 15/16 (1967), 263–77; Portugal: A. Reis, "List of Old Globes in Portugal," *Der Globusfreund*, 35/37 (1987), 249–54; Soviet Union: T.P. Matvejeva, "Alte Globen in der Sowjetunion," *Der Globusfreund*, 21/23 (1973), 226–33; USA: E.L. Yonge, *A catalogue of early globes* (New York: American Geogr. Society, 1968).
11. Luisa Martín-Merás started compiling a catalog of old globes in Spain in the 1990s, as acknowledged in L. Martín-Merás, "Los globos de la familia Blaeu en España," *Boletín de la Real Academia de la Historia*, 197 (2000), 497–512. However, the project was never completed or published. I am thankful to Luisa Martín-Merás for kindly sharing her ongoing list of globes with me; the 67 globes from public collections on her list were already part of the catalog published here, which includes up to 199 pieces.
12. Martín-Merás, *op. cit.* (Note 11).
13. Z. Navarro Abreu et al., "Preservando el patrimonio de la RSEAPT: Las esferas terrestre y celeste del legado de Nava," *Cartas diferentes. Revista canaria de patrimonio documental*, 15 (2019), 257–70.
14. M. Brañas et al., "Restauración de un globo terráqueo inglés del siglo XVIII en el Museo de América," *Anales del Museo de América* 11 (2003), 253–68.
15. A. Serrano, "Los globos celestes y terrestres," in *Investigación, conservación y restauración de materiales cartográficos y objetos cartográficos* (Madrid: Ministerio de Educación, Cultura y Deporte, 2010), pp. 81–96.
16. A. Hernando, "Die Herstellung von Erd- und Himmelsgloben in Spanien," *Der Globusfreund*, 59/60 (2014), 162–202.
17. https://ibercarto.ign.es
18. https://directoriomuseos.mcu.es
19. In the catalog, we distinguish between planetaria, where each planet (or moon) has to be moved individually by hand (similar to a heliocentric armillary sphere, but without full rings), and orreries, which are mechanical models of the solar system driven by a clockwork mechanism (i.e. a series of gears connect the various elements of the model).
20. Orreries or planetaria limited to the earth-sun, earth-moon, or earth-moon-sun system.
21. Purchased in 2015, currently undergoing a detailed study before restoration by the Instituto del Patrimonio Cultural de España.
22. Martín-Merás, *op. cit*. (Note 11).
23. The latter was purchased in 2002 and, thus, not yet listed in Martín-Merás, *op. cit*. (Note 11).
24. Sumira, *op. cit*. (Note 4), pp. 23–4.
25. Yonge, *op. cit.* (Note 10).

26. Additionally, the 46 cm terrestrial and celestial globes by George Adams from Museo Nacional de Ciencia y Tecnología and Palacio del Senado likely formed a pair, as well as the 63 cm celestial-terrestrial Delamarche globes from the University and Escuela de Estudios Árabes in Granada.
27. The cartography of the globe from the Morgan Library demonstrates that it was based on the map by Girolamo da Verrazzano in 1529, based on the explorations of Giovanni da Verrazzano that same decade; Stevenson, *op. cit.* (Note 9), pp. 105–6. This is particularly shown by the toponym of “Verrazana” for North America, as well as the appearance of the so-called sea of Verrazzano. The globe from Madrid also shows the inscription “VERRAZANA”, but mutilated as “ERRA□ANA”.
28. According to H. Harrisse, “Un nouveau globe Verrazanien,” *Revue de Géographie*, 37 (1895), pp. 175–7, also cited by Stevenson, *op. cit.* (Note 9), pp. 105–6, another 14 cm gilded globe signed “Robertus de Bailly 1530” was held at the Bibliothèque Nationale de France at the end of the 19th century, composed of “two parts rather insecurely joined on the line of the equator”. However, the Bibliothèque Nationale de France does not hold such a globe nowadays (it only preserves the modern printed gores of the globe from the Morgan Library, GE D-8144).
29. J. Siebold, “Bailly’s Globe [Verrazano Globe],” in *Renaissance Maps* (2021). Available at: https://www.myoldmaps.com/renaissance-maps-1490-1800/351-the-bailly-globe/ (accessed 10 January 2025).
30. J. Camón Aznar, *Guía del Museo Lázaro Galdiano* (Madrid: Fundación Lázaro Galdiano, 1973).
31. J.M. Cruz Valdovinos, *Platería en la Fundación Lázaro Galdiano* (Madrid: Fundación Lázaro Galdiano, 2000).
32. A. Drahos, *L'astronomie dans l'art: de la Renaissance à nos jours* (Paris: Citadelles & Mazenod, 2014).
33. Hernando, *op. cit.* (Note 16).
34. C. Líter Mayayo, *Biblioteca Nacional de España, 300 años haciendo historia* (Madrid: Biblioteca Nacional de España, 2011), pp. 174–5.
35. Hernando, *op. cit.* (Note 16).
36. Líter Mayayo, *op. cit.* (Note 34), pp. 174–5.
37. C. Líter Mayayo and F. Sanchis Ballester, *La obra de Tomás López : imagen cartográfica del siglo XVIII* (Madrid: Biblioteca Nacional, 2002).
38. F. Patier Torres, *La biblioteca de Tomás López: seguida de la relación de los mapas impresos, con sus cobres, y de los libros del caudal de venta que quedaron a su fallecimiento en Madrid en 1802* (Madrid: El Museo Universal, 1992).
39. One at the Maritime Museum in Greenwhich and another one in the private collection of the late Rudolf Schmidt. E. Dekker, *Globes at Greenwich* (Oxford: Oxford University Press, 1999). H. Wohlschläger, *The Rudolf Schmidt Globe Collection* (Vienna: International Coronelli Society for the Study of Globes, 2024).
40. Líter Mayayo, *op. cit.* (Note 34), pp. 72–3.
41. Dossier archival reference from Biblioteca Nacional: BNE-A RT/2098/028 and BNE-A BN 2861/013.
42. H. Richard, *Les Globes de Coronelli* (Paris: Bibliothèque national de France / Seuil, 2006).
43. T. Horst “Zur Entstehung der Manuskriptgloben des Heinrich Arboreus und des Philipp Apian,” *Der Globusfreund*, 57/58 (2011), 111–27.
44. Hernando, *op. cit.* (Note 16).
45. C. Hofmann, “Les globes de l'abbé Nollet,” in: B. Saule and C. Arminjon (eds.), *Sciences & curiosités à la cour de Versailles* (Paris: Réunion des Musées nationaux, 2010), pp.192–3.
46. The internal records from Museo de América show that this pair of globes by Dudley Adams comes from Biblioteca Nacional, on permanent loan since the foundation of the museum.
47. We note that, while Líter Mayayo, *op. cit.* (Note 34), pp. 72–3, lists a diameter of 75 cm for the blue celestial globe at the National Library, the diameter of the sphere measured for this work agrees with the 68 cm of the globes by Dudley Adams at Museo de América.
48. Líter Mayayo, *op. cit.* (Note 34), pp. 72–3.
49. E. La Parra López and E. Larriba, *Memorias de Manuel Godoy* (San Vicente del Raspeig: Publicaciones de la Universidad de Alicante, 2008).
50. M. López Arroyo, *El Real Observatorio Astronómico de Madrid (1785-1975)* (Madrid: Dirección General del Instituto Geográfico Nacional, 2004). J. Tinoco, *Apuntes para la historia del Observatorio de Madrid* (Madrid: Instituto Geográfico y Catastral, 1951).
51. La Parra López and Larriba, *op. cit.* (Note 49).
52. A. Aranda Huete, *La medida del tiempo: relojes de reyes en la corte española del siglo XVIII* (Madrid: Patrimonio Nacional, 2011), pp. 445–6.
53. Instituto Geográfico Nacional, *Instrumentos históricos del Instituto Geográfico Nacional* (Madrid: Centro Nacional de Información Geográfica, 2002). López Arroyo, *op. cit.* (Note 50).
54. López Arroyo, *op. cit.* (Note 50).
55. The inventory of instruments from the Observatory carried out by José Ramón de Ibarra in 1816 lists both armillary spheres, as well as the lunar globe by Russell, confirming that they were indeed associated with the Observatory since shortly after its foundation in 1790. The inventory also lists an English celestial-terrestrial globe pair by William Bardin (20 inch diameter), which now seems to be lost. José Ramón de Ibarra, *Lista de los Instrumentos del Observatorio Astronómico*. Archivo del Museo Nacional de Ciencias Naturales. Fondo Observatorio Astronómico. Legajo 1, carpeta 2.ª.

56. Two are preserved at Real Sociedad Económica Matritense de Amigos del País and one is held at Museo de Historia de Madrid (published under "Catálogo de la exposición Madrid hasta 1875. Testimonios de su historia", Museo Municipal de Madrid, 1979. Catalog number 1109).

57. One of them is preserved at Museo Nacional de Ciencia y Tecnología, on loan from IES San Isidro, and which could have served as a direct model.

58. C. Ovelar, "A Pair of Spanish Celestial and Terrestrial Globes from the Second Half of the Nineteenth Century: History, Singularity, and Restoration," *Globe Studies*, 69 (2025), 104–8.

59. S. Falk, "Ein Spanischer Globus: Entstehungsgeschichte und Interpretation," *Der Globusfreund*, 59/60 (2014), 142–61.

60. The provenance of this globe by Benjamín Tena is a donation from a private collection, and at least one more globe of this kind remains in private collections.

61. Amelia Aranda Huete, private communication.

62. J.R. Colón de Carvajal, *Catálogo de relojes* (Madrid: Patrimonio Nacional, 1987), pp. 142–3.

63. Aranda Huete, *op. cit.* (Note 52), p. 335.

64. C. Seguin, "Planétaire XVIIIe signé: Lemaire, 1788," *Horlogerie Ancienne* (2010), 85–96.

65. C. Seguin, "Le temps: La mesure du temps, du Big Bang à la pendule de cheminée," *Académie des Sciences et Lettres de Montpellier* (2015), 281–93.

66. Seguin, *op. cit.* (Note 65).

67. Seguin, *op. cit.* (Note 65).

68. IES Padre Suárez (Granada), IES Isabel de Castilla (Ávila), Colegio Público San Ildefonso (Madrid), and IES Histórico Alfonso VIII (Cuenca), the latter currently on loan at Museo de Ciencias Castilla-La Mancha. The one from Museo Marítimo del Cantábrico comes from the Escuela de Náutica de Santander.

69. Another example can be found at the Musée Joseph-Denais, in Beaufort-en-Anjou.

70. Some of them have a label indicating names such as "Aramburo H$^{os}$ / Calle del Príncipe 12, Madrid" (IES Histórico Alfonso VIII, Cuenca) or "Grasselli, Óptico Real de S.M." (IES Isabel de Castilla, Ávila). These are, in all likelihood, the names of the local companies which imported the pieces from Bertaux and sold them within Spain.

71. A.J. Turner, "Vitreous Globe: An 18th-century Novelty," in *Bulletin of the Scientific Instrument Society*, 47 (1995), 8–11.

72. Ovelar, *op. cit.* (Note 58). Hernando, *op. cit.* (Note 16), pp. 193–6.

73. Ovelar, *op. cit.* (Note 58). Hernando, *op. cit.* (Note 16), pp. 193–6.

74. Restaured in 2020: University of Granada (2020), "La restauración de un globo celeste de A. Delamarche." Available at: https://patrimonio.ugr.es/restauracion/la-restauracion-de-un-globo-celeste-de-a-delamarche (accessed 3 November 2025).

75. Sumira, *op. cit.* (Note 4), pp. 178–81. D. Smith, "The Cary family", in *The Map Collector,* 43 (Tring: Map Collector Publications, 1988), pp. 40–7. G. Clifton, *Directory of British scientific instrument makers, 1550-1851* (London: Zwemmer, 1995), p. 51.

76. E. Dekker and P. van der Krogt, *Globes from the Western World* (London: Zwemmer, 1993), p. 122. Smith, *op. cit.* (Note 75), pp. 40–7. Clifton, *op. cit.* (Note 75), p. 51.

77. Dekker & van der Krogt, *op. cit.* (Note 76), p. 122. Smith, *op. cit.* (Note 75), pp. 40–7. Clifton, *op. cit.* (Note 75), p. 51.

78. Sumira, *op. cit.* (Note 4), pp. 178–81. Smith, *op. cit.* (Note 75), pp. 40–7. Clifton, *op. cit.* (Note 75), p. 51.

79. Ovelar, *op. cit.* (Note 58). Hernando, *op. cit.* (Note 16), pp. 193–6.

80. Ovelar, *op. cit.* (Note 58). Hernando, *op. cit.* (Note 16), pp. 193–6.

81. Ovelar, *op. cit.* (Note 58). Hernando, *op. cit.* (Note 16), pp. 193–6.

82. Even though the date 1740 is written on the celestial globe, Dekker, *op. cit.* (Note 39), argues for a possible earlier production around 1730. Clifton, *op. cit.* (Note 75), p. 248. Dekker, *op. cit.* (Note 39), pp. 120–4, 493–5.

83. Even though the date 1740 is written on the celestial globe, Dekker, *op. cit.* (Note 39), argues for a possible earlier production around 1730. Clifton, *op. cit.* (Note 75), p. 248. Dekker, *op. cit.* (Note 39), pp. 120–4, 491–3.

84. Smith, *op. cit.* (Note 75), pp. 40–7. Clifton, *op. cit.* (Note 75), p. 51.

85. Navarro Abreu et al., *op. cit.* (Note 13). Tom Lamb, *The World in Your Hands: An Exhibition of Globes and Planetaria from the Collection of Rudolf Schmidt* (London: Christie's, 2005), p. 64. Dekker & van der Krogt, *op. cit.* (Note 76), p. 123. Smith, *op. cit.* (Note 75), pp. 40–7. Clifton, *op. cit.* (Note 75), p. 51.

86. Navarro Abreu et al., *op. cit.* (Note 13). Lamb, *op. cit.* (Note 85), p. 64. Smith, *op. cit.* (Note 75), pp. 40–7. Dekker, *op. cit.* (Note 39), pp. 296–8. Clifton, *op. cit.* (Note 75), p. 51.

87. Millburn, *op. cit.* (Note 6), pp. 157–270. Clifton, *op. cit.* (Note 75), pp. 2–3.

88. Van der Krogt, *op. cit.* (Note 5), pp. 140–60, 509–24. Martín-Merás, *op. cit.* (Note 11). E. Dahl, *Sphaerae Mundi: Early Globes at the Stewart Museum* (Montreal: Septentrion/McGill-Queen's University Press, 2000), pp. 73–4. Dekker & van der Krogt, *op. cit.* (Note 76), pp. 50–2. P.C.J. Van der Krogt, *The Most Magnificent and Largest Globes of Blaeu, the World's Greatest Globe Maker* (Leiden: Hes & de Graaf, 2001), pp. 12–7. Dekker, *op. cit.* (Note 39), pp. 290–2.
89. Van der Krogt, *op. cit.* (Note 5), pp. 140–60, 509–24. Martín-Merás, *op. cit.* (Note 11). Dahl, *op. cit.* (Note 88), pp. 73–4. Dekker & van der Krogt, *op. cit.* (Note 76), pp. 50–2. Van der Krogt, *op. cit.* (Note 88), pp. 1–11. Dekker, *op. cit.* (Note 39), pp. 287–90.
90. M.S. Pedley, *Bel et utile. The work of the Robert de Vaugondy Family of Mapmakers* (Tring: Map Collector Publications, 1992). Dekker & van der Krogt, *op. cit.* (Note 76), pp. 74–6. Dahl, *op. cit.* (Note 88), pp. 167–73.
91. Clifton, *op. cit.* (Note 75), p. 248.
92. Serrano, *op. cit.* (Note 15), pp. 89–96. Dekker & van der Krogt, *op. cit.* (Note 76), pp. 110–1. Millburn, *op. cit.* (Note 6), pp. 271–306. Clifton, *op. cit.* (Note 75), p. 2.
93. Hernando, *op. cit.* (Note 16), pp. 189-93.
94. Hernando, *op. cit.* (Note 16), pp. 189-93.
95. Hernando, *op. cit.* (Note 16), pp. 189-93.
96. Hernando, *op. cit.* (Note 16), pp. 189-93.
97. Dated 1700, published ca. 1707, Van der Krogt, *op. cit.* (Note 5), pp. 310–30, 550–4. Dekker, *op. cit.* (Note 39), pp. 511–2.
98. Dated 1700, published ca. 1711, Van der Krogt, *op. cit.* (Note 5), pp. 310–30, 550–4.
99. M.S. Pedley, *Bel et utile. The work of the Robert de Vaugondy Family of Mapmakers* (Tring: Map Collector Publications, 1992). Dekker & van der Krogt, *op. cit.* (Note 76), pp. 74–6. Dahl, *op. cit.* (Note 88), pp. 167–73.
100. M.S. Pedley, *Bel et utile. The work of the Robert de Vaugondy Family of Mapmakers* (Tring: Map Collector Publications, 1992). Dekker & van der Krogt, *op. cit.* (Note 76), pp. 74–6. Dahl, *op. cit.* (Note 88), pp. 167–73.
101. Ovelar, *op. cit.* (Note 58). Hernando, *op. cit.* (Note 16), pp. 193–6.
102. Dahl, *op. cit.* (Note 88), pp. 178–83. Dekker, *op. cit.* (Note 39), pp. 322–4.
103. Dahl, *op. cit.* (Note 88), pp. 178–83. Dekker, *op. cit.* (Note 39), pp. 324–5.
104. Wohlschläger, *op. cit.* (Note 39), p. 290.
105. Hernando, *op. cit.* (Note 16), pp. 189-93.
106. Dekker & van der Krogt, *op. cit.* (Note 76), p. 122. Smith, *op. cit.* (Note 75), pp. 40–7. Clifton, *op. cit.* (Note 75), p. 51.
107. Dekker & van der Krogt, *op. cit.* (Note 76), p. 122. Smith, *op. cit.* (Note 75), pp. 40–7. Clifton, *op. cit.* (Note 75), p. 51.
108. Hernando, *op. cit.* (Note 16), pp. 189-93.
109. Wohlschläger, *op. cit.* (Note 39), p. 295.
110. Van der Krogt, *op. cit.* (Note 5), pp. 310–30, 555–8. Wohlschläger, *op. cit.* (Note 39), pp. 97–8.
111. Dahl, *op. cit.* (Note 88), pp. 178–83.
112. Wohlschläger, *op. cit.* (Note 39), pp. 145–8.
113. Hernando, *op. cit.* (Note 16), pp. 189-93.
114. Hernando, *op. cit.* (Note 16), pp. 189-93.
115. Van der Krogt, *op. cit.* (Note 5), pp. 140–60, 509–24. Martín-Merás, *op. cit.* (Note 11). Dahl, *op. cit.* (Note 88), pp. 73–4. Dekker & van der Krogt, *op. cit.* (Note 76), pp. 50–2. Van der Krogt, *op. cit.* (Note 88), pp. 12–7. Dekker, *op. cit.* (Note 39), pp. 290–2.
116. Van der Krogt, *op. cit.* (Note 5), pp. 140–60, 509–24. Martín-Merás, *op. cit.* (Note 11). Dahl, *op. cit.* (Note 88), pp. 73–4. Dekker & van der Krogt, *op. cit.* (Note 76), pp. 50–2. Van der Krogt, *op. cit.* (Note 88), pp. 1–11. Dekker, *op. cit.* (Note 39), pp. 287–90.
117. Hernando, *op. cit.* (Note 16), pp. 189-93.
118. Same as held at Whipple Museum, Cambridge; see Falk, *op. cit.* (Note 59).
119. Lamb, *op. cit.* (Note 85), p. 76. Wohlschläger, *op. cit.* (Note 39), pp. 254–5. Dekker, *op. cit.* (Note 39), pp. 393–4. Clifton, *op. cit.* (Note 75), p. 162.
120. Wohlschläger, *op. cit.* (Note 39), p. 151.
121. Ovelar, *op. cit.* (Note 58). Hernando, *op. cit.* (Note 16), pp. 193–6.
122. Ovelar, *op. cit.* (Note 58). Hernando, *op. cit.* (Note 16), pp. 193–6.
123. Wohlschläger, *op. cit.* (Note 39), p. 151.
124. Dekker & van der Krogt, *op. cit.* (Note 76), p. 109. Sumira, *op. cit.* (Note 4), pp. 88–9. Clifton, *op. cit.* (Note 75), pp. 194–5.
125. Dekker, *op. cit.* (Note 39), pp. 120–4, 491–3. Clifton, *op. cit.* (Note 75), p. 248.
126. Dekker, *op. cit.* (Note 39), pp. 120–4, 493–5. Clifton, *op. cit.* (Note 75), p. 248.
127. Millburn, *op. cit.* (Note 6), pp. 107–32. Clifton, *op. cit.* (Note 75), p. 2.

128. Millburn, *op. cit.* (Note 6), pp. 107–32. Clifton, *op. cit.* (Note 75), p. 2.
129. Sumira, *op. cit.* (Note 4), pp. 150–3. Millburn, *op. cit.* (Note 6), pp. 107–32. Clifton, *op. cit.* (Note 75), p. 2.
130. Dekker, *op. cit*. (Note 39), pp. 128–9. Clifton, *op. cit.* (Note 75), p. 200.
131. Dekker & van der Krogt, *op. cit*. (Note 76), pp. 118–9. Clifton, *op. cit.* (Note 75), p. 200.
132. Hernando, *op. cit.* (Note 16), pp. 189-93.
133. H. King and J. R. Millburn: *Geared to the stars: the evolution of planetariums, orreries, and astronomical clocks* (Toronto, Buffalo : University of Toronto Press, 1978), pp. 304–5.
134. Wohlschläger, *op. cit*. (Note 39), p. 295.
135. Aranda Huete, *op. cit.* (Note 52), p. 336. Colón de Carvajal, *op. cit.* (Note 62), pp. 144–5. H.L. Tardy, *La pendule française dans le monde* (Paris: Tardy, 1981), p. 544. King and Millburn, *op. cit.* (Note 133), pp. 305–6.
136. Líter Mayayo, *op. cit.* (Note 34), pp. 174–5. Hernando, *op. cit.* (Note 16), pp. 183-7.
137. Líter Mayayo, *op. cit.* (Note 34), pp. 72–3.
138. Contains a celestial globe made of glass. Aranda Huete, *op. cit.* (Note 52), pp. 30–1, 211–3. Colón de Carvajal, *op. cit.* (Note 62), pp. 12. King and Millburn, *op. cit.* (Note 133), pp. 245–6.
139. Millburn, *op. cit.* (Note 6), pp. 271–306. Clifton, *op. cit.* (Note 75), p. 2.
140. Dekker & van der Krogt, *op. cit*. (Note 76), pp. 118–9. Smith, *op. cit.* (Note 75), pp. 40–7. Wohlschläger, *op. cit*. (Note 39), pp. 249–51. Clifton, *op. cit.* (Note 75), p. 51.
141. Lamb, *op. cit.* (Note 85), p. 65. Dekker & van der Krogt, *op. cit*. (Note 76), pp. 118–9. Smith, *op. cit.* (Note 75), pp. 40–7. Wohlschläger, *op. cit*. (Note 39), pp. 249–51. Clifton, *op. cit.* (Note 75), p. 51.
142. Dekker & van der Krogt, *op. cit*. (Note 76), p. 126.
143. Dekker & van der Krogt, *op. cit*. (Note 76), p. 126.
144. Clifton, *op. cit.* (Note 75), p. 92.
145. Van der Krogt, *op. cit*. (Note 5), pp. 392–3.
146. We list the date of first publication, but this globe continued to be produced without change for some decades afterward. Van der Krogt, *op. cit*. (Note 5), pp. 57–69, 413–15. Dekker, *op. cit*. (Note 39), pp. 91–5, 413–5. Dekker & van der Krogt, *op. cit*. (Note 76), pp. 31–7. Sumira, *op. cit.* (Note 4), pp. 58–61.
147. We list the date of first publication, but this globe continued to be produced without change for some decades afterward. Van der Krogt, *op. cit*. (Note 5), pp. 57–69, 413–15. Dekker & van der Krogt, *op. cit*. (Note 76), pp. 31–7. Sumira, *op. cit.* (Note 4), pp. 54–7. Dekker, *op. cit*. (Note 39), pp. 91–5, 412–3.
148. Van der Krogt, *op. cit*. (Note 5), pp. 140–60, 488–97. Martín-Merás, *op. cit*. (Note 11). Wohlschläger, *op. cit*. (Note 39), pp. 90–1. Dekker, *op. cit*. (Note 39), pp. 285–7.
149. Van der Krogt, *op. cit*. (Note 5), pp. 140–60, 509–24. Martín-Merás, *op. cit*. (Note 11). Dahl, *op. cit.* (Note 88), pp. 73–4. Dekker & van der Krogt, *op. cit*. (Note 76), pp. 50–2. Van der Krogt, *op. cit*. (Note 88), pp. 12–7. Dekker, *op. cit*. (Note 39), pp. 290–2.
150. Van der Krogt, *op. cit*. (Note 5), pp. 140–60, 509–24. Martín-Merás, *op. cit*. (Note 11). Dahl, *op. cit.* (Note 88), pp. 73–4. Dekker & van der Krogt, *op. cit*. (Note 76), pp. 50–2. Van der Krogt, *op. cit*. (Note 88), pp. 1–11. Dekker, *op. cit*. (Note 39), pp. 287–90.
151. Dahl, *op. cit.* (Note 88), pp. 154–61. Dekker & van der Krogt, *op. cit*. (Note 76), pp. 72–3. Dekker, *op. cit*. (Note 39), pp. 325–7.
152. H. Gaab, *Johann Gabriel Doppelmayr (1677-1750) : sein Leben, seine Schriften, seine Karten, seine Globen* (Leipzig : AVA - Akademische Verlagsanstalt, 2023). Wohlschläger, *op. cit*. (Note 39), pp. 176-8. Dekker, *op. cit*. (Note 39), pp. 328–9.
153. M. Brañas et al., *op. cit.* (Note 14). Millburn, *op. cit.* (Note 6), pp. 271–306. Clifton, *op. cit.* (Note 75), p. 2.
154. M. Brañas et al., *op. cit.* (Note 14). Millburn, *op. cit.* (Note 6), pp. 271–306. Clifton, *op. cit.* (Note 75), p. 2.
155. Clifton, *op. cit.* (Note 75), p. 176.
156. Clifton, *op. cit.* (Note 75), p. 176.
157. "Catálogo de la exposición Madrid hasta 1875. Testimonios de su historia" (Madrid: Museo Municipal de Madrid, 1979). Catalog number 1109.
158. Siebold, *op. cit.* (Note 29). Camón Aznar, *op. cit.* (Note 30). Cruz Valdovinos, *op. cit.* (Note 31).
159. Van der Krogt, *op. cit*. (Note 5), pp. 140–60, 488–97. Martín-Merás, *op. cit*. (Note 11). Dekker, *op. cit*. (Note 39), pp. 285–7.
160. King and Millburn, *op. cit.* (Note 133), p. 210. Dekker, *op. cit.* (Note 39), pp. 316–9. Clifton, *op. cit.* (Note 75), p. 125–6.
161. Dekker, *op. cit.* (Note 39), pp. 112–7. Dahl, *op. cit.* (Note 88), pp. 131–4. Sumira, *op. cit.* (Note 4), pp. 90–1. Hofmann and Nawrocki, *op. cit*. (Note 1), pp. 180–3.
162. Sumira, *op. cit.* (Note 4), pp. 92–5.
163. Wohlschläger, *op. cit*. (Note 39), pp. 183–4.

164. Smith, *op. cit.* (Note 75), pp. 40–7. Dekker, *op. cit*. (Note 39), p. 294. Clifton, *op. cit.* (Note 75), p. 51.
165. Dahl, *op. cit.* (Note 88), pp. 178–83.
166. Sumira, *op. cit.* (Note 4), pp. 144–9. Millburn, *op. cit.* (Note 6), pp. 107–32. Clifton, *op. cit.* (Note 75), p. 2.
167. The so-called "Atlas" clock likely represents Hercules. Manuscript blue celestial globe with orrery inside. Aranda Huete, *op. cit.* (Note 52), p. 335. Colón de Carvajal, *op. cit.* (Note 62), pp. 142–3. Tardy, *op. cit.* (Note 135), p. 338. King and Millburn, *op. cit.* (Note 133), pp. 286–90. Seguin, *op. cit.* (Note 64). Seguin, *op. cit.* (Note 65).
168. Contains a manuscript blue celestial globe. Aranda Huete, *op. cit*. (Note 52), pp. 279–81. Colón de Carvajal, *op. cit.* (Note 62), pp. 110.
169. Contains a small terrestrial globe. Aranda Huete, *op. cit*. (Note 52), p. 338. Colón de Carvajal, *op. cit.* (Note 62), pp. 208. Tardy, *op. cit.* (Note 135), p. 276.
170. Van der Krogt, *op. cit*. (Note 5), pp. 140–60, 488–97. Martín-Merás, *op. cit*. (Note 11). Dekker, *op. cit*. (Note 39), pp. 285–7.
171. Van der Krogt, *op. cit*. (Note 5), pp. 140–60, 488–97. Martín-Merás, *op. cit*. (Note 11). Dekker, *op. cit*. (Note 39), pp. 284–5, 308–9.
172. Dekker, *op. cit*. (Note 39), pp. 125–7, 484–6. Sumira, *op. cit.* (Note 4), pp. 174–7. King and Millburn, *op. cit.* (Note 133), pp. 313–4. Clifton, *op. cit.* (Note 75), p. 240. López Arroyo, *op. cit.* (Note 50).
173. Attributed to the workshop of the Royal Observatory of Madrid. López Arroyo, *op. cit*. (Note 50). M. Comes, *op. cit*. (Note 175), pp. 212–3.
174. Attributed to the workshop of the Royal Observatory of Madrid. López Arroyo, *op. cit*. (Note 50). M. Comes, *op. cit*. (Note 175), pp. 212–3.
175. Hernando, *op. cit.* (Note 16), pp. 197-200.
176. M. Comes: *Historia de la esfera armilar : su desarrollo en las diferentes culturas* (Barcelona: Universitat de Barcelona, 2012), pp. 259–63.
177. Van der Krogt, *op. cit*. (Note 5), pp. 140–60, 509–24. Martín-Merás, *op. cit*. (Note 11). Dahl, *op. cit.* (Note 88), pp. 73–4. Dekker & van der Krogt, *op. cit*. (Note 76), pp. 50–2. Van der Krogt, *op. cit*. (Note 88), pp. 12–7. Dekker, *op. cit*. (Note 39), pp. 290–2.
178. Van der Krogt, *op. cit*. (Note 5), pp. 140–60, 509–24. Martín-Merás, *op. cit*. (Note 11). Dahl, *op. cit.* (Note 88), pp. 73–4. Dekker & van der Krogt, *op. cit*. (Note 76), pp. 50–2. Van der Krogt, *op. cit*. (Note 88), pp. 1–11. Dekker, *op. cit*. (Note 39), pp. 287–90.
179. Clifton, *op. cit.* (Note 75), p. 223.
180. Clifton, *op. cit.* (Note 75), p. 223.
181. Van der Krogt, *op. cit*. (Note 5), pp. 392–3.